\documentclass[12pt,preprint]{aastex} 

\shorttitle{Cosmological Impact of Classical Doubles}
\shortauthors{Barai \& Wiita}

\slugcomment{Submitted to The Astrophysical Journal} 

\begin{document} 


\title{Testing Models of Radio Galaxy Evolution and the Cosmological Impact of FR II Radio Galaxies}


\author{Paramita Barai\altaffilmark{1}, Paul J. Wiita} 
\affil{Department of Physics \& Astronomy, 
Georgia State University, 
P.\ O.\ Box 4106, Atlanta, GA 30302-4106, USA} 
\email{pabar56@phy.ulaval.ca, wiita@chara.gsu.edu}
\altaffiltext{1}
{current address: D{\'e}partement de physique, de g{\'e}nie physique et d'optique, 
Universit{\'e} Laval, Qu{\'e}bec, QC G1K 7P4, Canada.} 



\begin{abstract} 
{ 
We investigate aspects of the cosmological evolution of 
FR II radio galaxies, focusing first on the 
abilities of models to match data for 
linear-sizes, radio powers, redshifts and spectral indices. 
Here we consider modifications to the theoretical models 
we had treated earlier, 
primarily by accounting for the growth of the radius of hotspots with source size. 
Better fits to the distributions of most of the data in three 
low-frequency surveys can be found with sensible choices of model parameters 
but no model yet considered gives a good match to 
all of the survey data simultaneously, nor does any do a good job of producing 
the spectral index distributions. 
The observational datasets are too small to completely discriminate among the models. 
We calculate the volume fraction of the ``relevant universe'' 
cumulatively occupied by the expanding radio galaxy lobes over the quasar era, 
when these powerful radio galaxies were much more common, 
and when they have been argued to play an important role in 
triggering galaxy formation and spreading magnetic fields and metals. 
We found the cumulative relevant volume filling factor of radio galaxies to be $\sim 5 \%$, so
we conclude that 
these impacts are smaller than previously estimated 
but that they are still significant. 
} 
\end{abstract} 





\section{Introduction} 
\noindent 
Classical double radio galaxies (RGs) with extended lobes,   
or the Fanaroff-Riley Class II (FR II) sources, 
constitute the more powerful RG population. 
The  {\it quasar era} (i.e., between redshifts $\simeq$ 1.5 and 3) 
is distinguishable by rises in the comoving luminosity densities 
of RGs, quasars and other powerful AGN 
\citep{dunlop90, jackson99, willott01, ueda03, grimes04, hopkins06}, 
as well as high star and galaxy formation rates 
\citep[e.g.,][]{lilly96, madau98, bouwens06, sawicki06a, sawicki06}. 
These similar evolutions suggest that powerful RGs can have substantial impacts 
on the formation, distribution and evolution 
of galaxies and large scale structures of the universe 
(e.g., \citealt{GKW01}, GKW01; \citealt{kronberg01}; 
\citealt{GKW03a}, GKW03a; \citealt{GKW03b}, GKW03b; \citealt*{GKWO}, GKWO; 
\citealt*{GKWB04}, GKWB04; \citealt{rawlings04, levine05, silk05}). 
Observational investigations show remarkable radio-optical alignment in high-$z$ RGs 
\citep[e.g.,][]{mcCarthy87, chambers88a, chambers88b, dey97, 
venemans04, venemans05, greve05, roderik05, zheng06, ajiki06}. 
Some studies attribute this to FR II RGs triggering 
extensive star formation in a multi-phase intergalactic medium (IGM)
(e.g., \citealt{begelman89, deYoung89, rees89, chokshi97}; GKW01). 
In addition, the expanding RG lobes could easily have 
infused significant magnetic fields into the IGM (GKW01, GKWO, GKWB). 
Similar conclusions were drawn from different lines of argument by 
\citet{kronberg01} and \citet{furlanetto01}. 
These radio sources born through the quasar era 
could also have contributed toward metal enhancement 
of their environments (GKW03b; GKWB04), 
since observations \citep[e.g.,][]{dietrich03, schaye03, aguirre04, shapley04, 
pieri06, prochaska06, tripp06} 
require an efficient mechanism for spreading metals widely 
at early cosmic epochs. 

A comprehensive study of the impact of RGs on 
various events in the cosmological history of the universe 
requires reliable quantitative estimates of the 
relevant volume filling fraction of RGs and their active lifetimes.
A prerequisite for a more accurate computation of this volume impacted
by radio lobes is a good model of the evolution of radio sources,
for both individual sources and as a function of redshift.
In a recent work \citet{rawlings04} 
agreed that RG lobes will penetrate much of the relevant universe,
but they argued that this may often shut off star formation
by expelling gas from protoclusters.
However, they assume a single phase medium, 
unlike our picture or that of \citet{rees89}, 
so this negative conclusion is not surprising. 
Many recent observational studies 
\citep[e.g.,][]{bohringer95, birzan04, choi04, odea04, reynolds05} 
show depressions in X-ray surface brightness coinciding with 
radio lobes, cavities and buoyant bubbles in clusters of galaxies;
these are clear signatures of the interactions of radio sources with their 
surrounding hot intracluster gas 
on scales of several tens to hundreds of kiloparsecs from the AGN.


Our aim is to probe in more detail the cosmological impact of 
RGs on the growth of structures. 
To do so, we must first develop 
an improved but essentially analytical model for the evolution of 
FR II RGs as they age individually 
and as their numbers vary with cosmological epoch. 
As our first step toward this goal we \citep[][BW]{BW06} 
compared three fairly sophisticated analytical models 
for the dynamical and lobe power evolution of FR II RGs, 
those given by \citet*[KDA]{KDA},
\citet*[BRW]{BRW} and \citet[MK]{MK}; 
the reader should have some familiarity with these papers or see 
\citet{barai06} for more details. 
The source linear-size evolution in the BRW and MK models
essentially follow the KDA prescription.
However the models differ in the ways the relativistic particles 
are injected from the jet to the lobe,
and in treatments of loss terms and particle transport.
So there are some significant differences in their predictions for
observed powers ($P$) as functions of source size ($D$) and redshift ($z$). 

The $[P$--$D]$ evolutionary tracks of model radio sources have been 
used in KDA, MK, and in \citet{machalski04a, machalski04b} 
as the main way to look for consistency between 
observational data and RG evolution models. 
The radio sky simulation prescription in 
\citet*[BRW]{BRW} adds new dimensions to the observed parameter space.
Deriving a RG redshift distribution function 
from the work of \citet{willott01} on the radio luminosity function (RLF), 
BRW prescribed a virtual RG survey technique. 
This involves generation of a huge initial population of sources over extended cosmic epoch
according to pre-defined distribution functions 
in redshift, jet power, source age, and orientation. 
These large numbers of sources then evolve through their 
individual lives where they interact with their environments and
undergo energy losses. 
Only a few simulated sources are 
detected in the virtual surveys when mathematical flux limits,
corresponding to observations, are imposed. 
We adopted observational samples from the redshift-complete subsamples: 
3CRR \citep*{laing83}, 6CE \citep*{rawlings01} and 7CRS \citep{mcGilchrist90}, 
of the Cambridge flux-limited radio catalogs; see Table~2 of BW. 
The models, relevant observations and the 
multi-dimensional Monte Carlo simulation methodology leading to the virtual surveys 
are discussed in detail in \S\S 2 -- 4 of BW, 
and given in more detail in \citet{barai06}. 
Using the virtual surveys 
and any radio lobe power evolution model (KDA, BRW or MK),
one can get $P$, $D$, $z$, and spectral index, $\alpha$
(with the convention $P_{\nu} \propto \nu^{-\alpha}$),
values for the 
simulated model radio sources detected in the pseudo-surveys.
The distributions of the characteristics of these simulated RGs 
can then be compared to observational data to test the model. 
In order to assess the 
success of a theoretical model, 
we perform several statistical tests 
and present the quantified results for the various model fits. 
In BW, our main conclusion 
is that none of the heretofore published models provides an adequate description of the data. 

As our next major step toward the goal of isolating good RG models, 
we have varied the radio lobe power evolution models in the present work. 
We consider modifications to the original models of BRW and MK, 
by incorporating a variable hotspot size growing with the source age. 
Here, 
we perform analogous simulation-based virtual surveys on the modified models, 
and present the corresponding statistical results. 
In \S\ref{sec:ModifiedModels} we describe the models: 
Modified BRW (MBRW), Modified MK (MMK), K2000 \citep{kaiser00} 
and the alternative radio luminosity function (RLF). 
We give the statistical test results of the model simulations in \S\ref{sec:ModelSimResults}, 
which we discuss in \S\ref{sec:model-compare-diss}. 

One important goal of this work has been to address the question of
what fraction of the relevant volume of the universe
(the volume containing most of the cosmic baryons)
did the radio lobes occupy during the quasar era (GKW01). 
The warm/hot intergalactic medium (WHIM) comprises the main repository of cosmic baryons
which can potentially collapse to form 
star clusters or galaxies \citep[e.g.,][]{cen99}. 
So the radio lobes need to penetrate a significant portion of this
``relevant volume of the universe'' occupied by the WHIM filaments 
in order to have a significant role in impacting star formation
and spreading magnetic fields and metals 
(e.g., GKWB04; BGKOW04 and references therein).
In \S\ref{sec:RelVol-Fill-Frac} we calculate the total volume filled by the RGs 
over the quasar era (when their population peaked) 
as a fraction of the relevant volume of the universe. 
Our conclusions are in \S\ref{sec:Conclusions}.

\section{Modified Radio Galaxy Evolution Models}
\label{sec:ModifiedModels}

\begin{figure} [!t]
\centering
\includegraphics[scale=0.4]{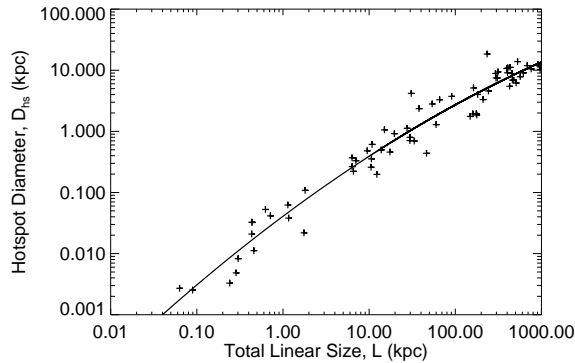}
\caption{Quadratic curve fit to the hotspot size vs.~linear size, 
$[D_{hs}$--$L]$ data of \citet{jeyakumar00} (\S\ref{sec:ModifiedModels}).}
\label{fig:DhsLfitJS}
\end{figure}

The major modification to the BRW and MK models involved allowing 
the hotspots to grow in size as a source ages and expands. 
The data used to make a sensible modification 
are taken from \citet[][hereafter JS00]{jeyakumar00}, 
who studied the dependence of sizes of hotspots on 
overall source sizes for a sample of FR II sources which included both 
compact steep spectrum and larger-sized sources 
spanning a projected source size range from about 50 pc to nearly 1 Mpc (Fig.~\ref{fig:DhsLfitJS}).
We parameterize the hotspot radius, 
\begin{equation}
r_{hs} = r_{hs_0} + f(L),   
\label{eq:r_hs-JS}
\end{equation}
where $r_{hs_0}$ is some normalizing initial hotspot radius,
and $f(L)$ is a power law expression of the total linear size $L$ of the source.
We chose $r_{hs_0}$ such that the hotspot of a source grows to
$r_{hs} = 2.5$ kpc when the total linear size is $L= 200$ kpc,
since these are reasonable averages of the actual values and
$r_{hs} = 2.5$ kpc was the constant value assumed by BRW.

The hotspot and source angular size data are
adopted from JS00. 
We follow these authors and find the average angular
hotspot size for each source. 
This is the geometric mean of each hotspot (major and minor axes) sizes and,
for those sources with hotspots detected on both sides,
the arithmetic average of their sizes. 
We convert the angular hotspot sizes and separations to corresponding 
projected linear sizes using the consensus cosmology \citep{spergel06}. 
Total linear sizes ($L$) are obtained by assuming an average angle
to the line of sight of $39.5 ^\circ$
(following KDA, as done for the $[P$--$D]$ tracks in BW \S5.1). 
The hotspots are much smaller than the total source sizes and
are assumed to be spherical,
so projection effects are negligible for them.

We performed various least-squares fits to the $\log(r_{hs})$ vs. $\log(L)$ data: 
a single straight line;
a single quadratic; 
two straight lines  with a break at $20$ kpc;  
two straight lines  with a break at $1$ kpc. 
Although all these fits are satisfactory, a quadratic fit to the data gave the lowest reduced $\chi^2$ and is shown in Fig.~\ref{fig:DhsLfitJS}. 
So the hotspot size was taken to be growing with the source size as 
\begin{equation}
\log (D_{hs}) = c_{hs_1} + y_1 \log L + y_2 \left( \log L \right)^2. 
\label{eq:log-D-hs} 
\end{equation} 
Then the hotspot radius was given by Eq.~(\ref{eq:r_hs-JS}) using $f(L) = D_{hs}/2$. 
The best fit values of the coefficients are 
$c_{hs_1} = -3.199, y_1 = 1.053$, and $y_2 = -0.0306$, 
which gave 
a reduced $\chi_r^2 = 0.2720$. 
\citet{perucho03} presented a dynamical model for compact-symmetric 
and FR II sources where the hotspot size grows self-similarly with linear size, 
and can get a fit which is in agreement with the results of \citet{jeyakumar00}.

In the modified BRW (MBRW) model the hotspot size (and area, $A_{hs} = \pi r_{hs}^2$)
rises according to Eqs.~(\ref{eq:r_hs-JS}) and (\ref{eq:log-D-hs}). 
It otherwise follows the prescription of the original
BRW model summarized in \S3.3 of BW. 
During a source's evolution several additional quantities
(which were fixed for a source when a constant hotspot size was assumed)
also varied with its age. 
These include the hotspot pressure, $p_{hs}$, 
the hotspot magnetic field, $B_{hs}$, 
the ``slow'' and ``fast'' break frequencies, $\nu_{bs}$ and $\nu_{bf}$, 
and the corresponding critical Lorentz factors, $\gamma_{bs}$ and $\gamma_{bf}$. 


The modified MK (MMK) model also incorporates the same rising hotspot size (and area).
In the MK model, the characteristic time, $t_0$, (Eq.~6 of MK) 
when the size of the head was comparable to the hotspot size,
depends on ``an initial'' hotspot area $A_{hs(t_0)}$, 
which we must now distinguish from the normal rising hotspot area $A_{hs}$.
We found this ``initial'' hotspot area using $r_{hs(t_0)} = 0.02$ kpc.
We chose this value as it gave the best 
results when we compared the statistics of 6 MMK simulation runs done
using $r_{hs(t_0)} = 0.01 - 0.06$ kpc, computed at intervals of 0.01 kpc.
So in the MMK model $t_0$ is 
\begin{equation}
t_0 = \left[ \frac{3 c_1^{2-\beta} c A_{hs(t_0)}}{\left(\Gamma_x+1\right) \left(5-\beta\right)^2} \right]^{1/a}
      \left( \frac{\rho_0 a_0^{\beta}}{Q_0} \right)^{3/\left(4+\beta\right)},
\end{equation}
with $A_{hs(t_0)} = \pi r_{hs(t_0)}^2 = \pi \left(0.02~{\rm kpc}\right)^2 $, 
$\rho_0$ is the central density of the ambient gas, 
$a_0$ is its scale length, $\beta$ is its radial density index, 
$Q_0$ is the jet power, $\Gamma_x$ is the adiabatic index of the external environment, 
$a$ is $\left( 4+\beta \right) / \left( 5-\beta \right)$, 
and $c_1$ is a model dependent (but weakly varying) constant 
(see BW for details). 
The MMK model otherwise follows the prescription of the 
original MK model as described in \S3.4 of BW.


\citet[][K2000]{kaiser00} proposed a modification
to the KDA \citep*{KDA} model, which essentially
follows the formulation in KDA but which has the following differences. 
The \citet{KA} and the KDA model considered a cylindrical geometry for the cocoon,
where the hotspot pressure $p_h$ drives the source expansion along the jet axis and
the expansion perpendicular to the axis is governed by the cocoon pressure $p_c$.
Hence it was assumed in the KDA model that
\begin{equation}
\frac{p_h}{p_c} = 4 R_T ^ 2 ,
\label{eq:ph-by-pc}
\end{equation}
where $R_T$ is the axial ratio, or the ratio of the length of the source
to the full width of a lobe half way down the jet. 
In subsequent studies of the shocked gas flow between the bow shock
and the cocoon, \citet{kaiser99b} empirically fitted
$\left(p_h/p_c\right)$ as functions of $\beta$ and $R_T$.
Such investigations appeared to show that the 
ratio in Eq.~(\ref{eq:ph-by-pc}) was an overestimate.
To correct this problem K2000 
claimed that a better empirical formula was given by
\begin{equation}
\frac{p_h}{p_c} = \left(2.14 - 0.52 \beta \right) R_T^{2.04-0.25\beta} .
\end{equation}


The source birth function over redshift,
\begin{equation}
 \rho(z) \propto {\rm exp}\left[- \frac{1}{2}
                      \left( \frac{z-z_0}{\sigma_z} \right)^{2}\right], 
\label{eq:rho-z} 
\end{equation}
is the common gaussian functional form of all the 
Radio Luminosity Functions (RLFs) we considered. 
A peak redshift of $z_0 = 2.2$, and a standard deviation of $\sigma_z = 0.6$, 
following \citet{willott01}, 
was adopted in the model simulations we performed in BW 
and in most of this paper. 
To explore the effect of different assumed redshift distributions we also 
considered extensively the RLF given by \citet*{grimes04}, which has 
$z_0 = 1.684$ and $\sigma_z = 0.447$, 
using the two-population generalised luminosity function 
from Table~5 of \citet{grimes04}. 

\section{Model Results} 
\label{sec:ModelSimResults} 

\subsection{$[P$--$D]$ Tracks} 
\label{sec:P-D-tracks} 

\begin{figure} [!t]
\centering
\includegraphics[scale=0.5]{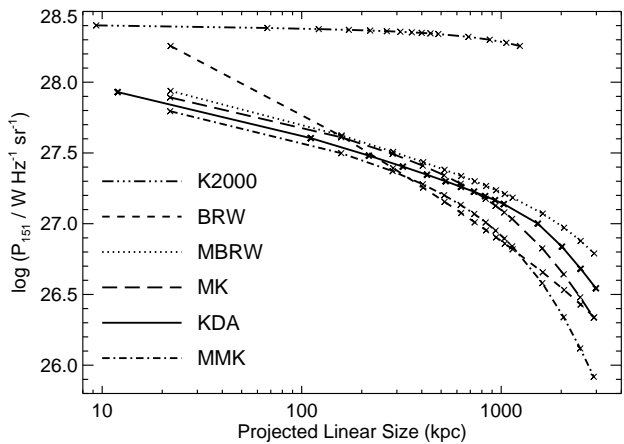}
\caption{$[P$--$D]$ tracks of a fiducial source having 
jet power, $Q_0 = 1.3 \times 10^{39}$ Watts, and at redshift, $z = 0.5$. 
The curves of different linestyles correspond to 
the tracks predicted by different models as labeled. 
The crosses on each track denote source lifetimes of 
1, 10, 20, 30, ..., 90, 100, 150, 200, 250, 300 Myr. 
}
\label{fig:PDtracksALLmod}
\end{figure}

The power ($P$) vs.\ linear-size ($D$), or, 
$[P$--$D]$ tracks of the MBRW, MMK, and K2000 models, 
as well as those of the original BRW, MK, and KDA models 
are shown in Fig.~\ref{fig:PDtracksALLmod}. 
The modified models follow the same general trends as do the 
original models described in \S5.1 of BW. 
The tracks are generated using the modified models
(described in \S\ref{sec:ModifiedModels}) 
with the default values of parameters for dynamical and power evolution 
from each of the original models (given in Table~1 of BW).
Each source was evolved at frequency $\nu=151$ MHz.
For this figure the total linear sizes were converted to the projected sizes
assuming an average viewing angle to the line of sight of
$39.5^{\circ}$ (following KDA). 

The rates of steepening of the tracks are 
significantly different in the three new models.
The MBRW track is less steep than the original BRW track. 
Among the three original models, KDA, BRW and MK,
BRW gave the worst fit to the data, when compared with respect to 
K-S statistical tests (BW). 
If the reason for this can be identified with the fact that BRW gave 
the steepest $[P$--$D]$ tracks (Fig.~1 of BW),
then this ``shallowing'' of the tracks in the MBRW model would imply
that the K-S statistical fits to the data
should be better for the MBRW model
(which is indeed true). 
The MMK track is slightly steeper than the original MK track, implying
less of a difference between their fits to the data. 
The K2000 track is much flatter than the 
corresponding typical tracks of any other model. 

The ``youth--redshift degeneracy'' \citep[e.g.,][]{blundell99}  
is clear in these $[P$--$D]$ tracks.
A high-power, high-redshift, source shows a
faster fall off in its specific $151$ MHz luminosity with time,
and can even fall below the limiting flux of a radio survey at a younger age,
as compared to a lower-power  source at a lower $z$.


\subsection{Preliminary Statistics: 1-D K-S Tests} 
\label{sec:prelim-1D-KS-test} 

We used 1-dimensional Kolmogorov-Smirnov (1-D K-S) statistics 
as a first quantitative test. 
Each of the distributions of key characteristics $[P, D, z, \alpha]$ 
of the radio sources detected in the 
simulated surveys were compared to those of the sources 
in the real radio surveys 3C, 6C, and 7C. 
The K-S probabilities, $\cal P$, that the two data sets
being compared are drawn from the same distribution function, were taken
to be figures of merit of each simulation.
To quantify the overall success of a model, 
we added the K-S probability statistic for comparisons of $P, D, z, \alpha$
(i.e., ${\cal P}(P)+{\cal P}(D)+{\cal P}(z)+{\cal P}(\alpha)$) for the three surveys,
weighting the statistic of a survey by the square-root
of the number of simulated sources detected in that survey, 
and denote this overall figure of merit as ${\cal P}_{[P, D, z, \alpha]}$. 
The second figure of merit we employ, denoted ${\cal P}_{[P, 2D, z, \alpha]}$,
adds the K-S statistic probabilities for $P$ and $z$ to twice those for
 $D$, i.e. ${\cal P}(P)+2{\cal P}(D)+{\cal P}(z)+{\cal P}(\alpha)$
using the same weighting method.
More details can be found in \S5.2.1 of BW. 
For each model some parameters were chosen as ``better'' 
(in providing 1-D K-S fits to the data); 
for these, further simulations and additional statistical tests 
(\S\ref{sec:2D-KS-test} and \ref{sec:corr-coeff-anal}) were done.

The procedures followed in doing the simulations and in presenting
the results are briefly discussed below;
for complete details see \S5 of BW. 
Table~\ref{tab:1DKS-MBRW} gives our results for the MBRW model; 
Table~\ref{tab:1DKS-MMK} gives those for the MMK model, and 
Table~\ref{tab:1DKS-K2000} those for the K2000 model. 
The tables follow the same format and pattern as the 
corresponding tables 3, 4 and 5 of BW for the original models, 
and should be compared to them. 
All the statistical test (K-S and correlations) 
results for the modified models include a 1 kpc cutoff of source size, i.e., 
the statistics are calculated by excluding sources with linear size, $D < 1$ kpc (see \S5.2.3 of BW). 

An initial ensemble, generated using the default parameters
from BRW for the RG population generation,
was evolved according to each of the modified models (MBRW, MMK and K2000). 
The sources in the simulated surveys 
(produced according to the prescription of \S4.3 of BW) 
were compared to the data samples of the 3C, 6C and 7C catalogs. 
We examined the 1-D K-S test statistics of the
first entry (the very first 3 rows) of 
Tables~\ref{tab:1DKS-MBRW}, \ref{tab:1DKS-MMK} and \ref{tab:1DKS-K2000} 
(modified model results)
and compared those to the first entries of Tables~4, 5 and 3 of BW, respectively 
(original model results). 
From this single comparison we find that
the MBRW model is significantly better 
(the combined 1-D K-S probabilities are $\sim 10 - 20$ times higher, 
which is $\sim 4$$\sigma$ better) than the original BRW model,
while the default MMK model is only slightly better than the original MK model. 
The K2000 model produces much worse K-S fits than did the original KDA model, 
and for the reasons discussed in \S\ref{sec:model-compare-diss} 
we did not explore the K2000 variation any further. 

In search of further improvements of the combined 1-D K-S statistics, 
we varied the beam power distribution 
function of the sources generated in the initial population
by checking steeper exponents, $x$, in the power-law of the
initial jet power distribution, 
where, $p(Q_0) \propto Q_0^{-x}$ between $Q_{min}$ and $Q_{max}$. 
For the MBRW model the overall statistics improved the most for $x=3.0$, 
while for the MMK model $x=2.6$ gave better fits. 
These values of $x$ 
were then used for the later simulations.

The initial population generated with $x=3$
(but otherwise using the BRW prescription),
was evolved according to the MBRW power evolution model. 
The corresponding 1-D K-S statistics are given in the third entry in Table~\ref{tab:1DKS-MBRW}, 
and show the improved fit (compared to the first entry of that table). 
To search for possible further improvements we varied the other parameters 
prescribing the power evolution in the models. 
Simulated surveys were constructed using the parameter listing
given in Table~\ref{tab:1DKS-MBRW} (each variation done one at a time) 
of the MBRW power evolution model.
The total 1-D K-S statistics, 
${\cal P}_{[P, D, z, \alpha]}$ and ${\cal P}_{[P, 2D, z, \alpha]}$, 
as seen from Table~\ref{tab:1DKS-MBRW}, 
are comparable to or better than the original BRW model results 
(Table~4 of BW). 

In order to find the best-fit maximum RG age, 
we performed simulation runs using initial populations with 
$x=3$ (for MBRW), and $x=2.6$ (for MMK), 
and then varied $T_{Max}$. 
The $T_{Max}$ that gave the highest mean values of the combined statistics
(${\cal P}_{[P, D, z, \alpha]}$ and ${\cal P}_{[P, 2D, z, \alpha]}$)
was chosen as the best maximum age at the considered $x$.
In the MBRW model the highest mean statistics
were seen at $T_{Max}$ = 300 Myr.
The MMK model performed its best at $T_{Max}$ = 150 Myr.
Hence we used initial populations with 
the above ``optimal'' $x$ and $T_{Max}$ pairs for each model in subsequent runs.
The 1-D K-S results of a subset of the first set of runs employing
these ``optimal'' initial populations and ages 
but varying some of the power evolution model parameters to alternate values 
are in Tables~\ref{tab:1DKS-MBRW} and \ref{tab:1DKS-MMK} for the MBRW and MMK models, respectively. 

Upon examining these preliminary results, 
only those cases that gave any improvement in statistics over the default case 
or were essentially as good as the default were considered further. 
For these parameter sets three more runs 
(making a total of four) were done using
the same large population but with different pseudo-random seeds,
yielding different ensembles of simluated samples.
Then the means and standard deviations of the
relevant 1-D K-S statistics were found.
Some ``2-change'' cases, i.e., models where two ``superior'' parameter variations
(those giving high 1-D K-S probabilities)
were simultaneously employed also were explored. 
As seen from the tables showing the individual 1-D K-S statistic probabilities 
(Tables~\ref{tab:1DKS-MBRW} and \ref{tab:1DKS-MMK}), 
often several of the 12 K-S probabilities for some cases give acceptable fits,
and a few would give very good fits, 
but it is difficult to find a single model where all are good fits. 
In other words, the modified models also do not give 
good {\it simultaneous} fits to the $[P, D, z, \alpha]$ data 
from all three of the radio surveys considered (3C, 6C and 7C).


The 1-D K-S statistical test results for the simulation runs 
of different models using the \citet{grimes04} RLF are given in 
Table~\ref{tab:1DKS-GrimesRLF}. 


\subsection{2-D K-S Tests}
\label{sec:2D-KS-test}

We performed additional statistical analyses on the modified models, so as
to make more robust comparisons both between them and with the original models.
This was done in a fashion similar to that of \S5.3 of BW for the original BRW, KDA and MK models. 
For each modified model 
the 1-D K-S best-fit parameter variation cases, 
i.e., those which gave the highest combined probability ${\cal P}_{[P, D, z, \alpha]}$, 
according to the results from the previous sub-section were selected. 

The 2-D K-S test results for both the default versions 
of the modified models and the parameter sets (denoted as ``varied'') 
giving the highest total 1-D K-S probability for each model, 
are given in Table~\ref{tab:2DKS-all}. 
The results are listed in a similar way as are the 1-D K-S statistics in previous tables, 
and are discussed below (\S\ref{sec:model-compare-AddStatTest}).

\subsection{Correlation Coefficients} 
\label{sec:corr-coeff-anal} 

Spearman partial rank correlation coefficients 
were calculated for those cases for which the 2-D K-S tests were done. 
We combined the $[P, D, z, \alpha]$ data from the 3 surveys: 3C, 6C and 7C III, 
for the actual observations and for the model simulations, 
and computed correlations between them 
(see \S5.3.2 of BW for the limitation to 7C III). 
This was mainly done in order to subdue the tight $[P$--$z]$ correlation 
present in a single flux-limited complete survey, 
and to thereby discover any correlations which exist between the other source characteristics. 

Table~\ref{tab:CorrCoeff-all} gives the four-variable 
Spearman partial rank correlation coefficients ($r_{PD, z\alpha}, r_{Pz, D\alpha}$, etc.) 
which were computed on the combined data and on the simulated results
from the modified models. 
We also examined the corresponding 2-variable and 3-variable correlations.
Probably the most important result is that the 2-variable correlation, $r_{PD}$, is always negative; 
however, when the 4-variable correlation between $P$ and $D$ is found, 
(i.e., with the effects of $z$ and $\alpha$ removed), 
a small positive correlation is seen between $P$ and $D$ 
(i.e., $r_{PD, z\alpha} > 0$). 
These are the same as trends seen in the $[P$--$D]$ correlations 
for the original models. 

\section{Discussion of Simulation Results} 
\label{sec:model-compare-diss} 

The K2000 model produces very flat $[P$--$D]$ tracks (Fig.~\ref{fig:DhsLfitJS}) 
and the 1-D K-S statistical results obtained in 
the multi-dimensional Monte Carlo simulations (Table~\ref{tab:1DKS-K2000}), 
are all very poor compared to any of the other models studied here or in BW. 
So we conclude that the K2000 model 
cannot well reproduce the trends of observed properties in the low frequency radio surveys.
The reason for such poor behavior of this model appears to be that the K2000 model was
designed specifically  to describe the environments and ages of
three local, and rather atypical, FR II sources.
As this model is biased toward describing special environments,
the parameters used in K2000 cannot really be applied globally
and hence it is not surprising that this model cannot explain the cosmological evolution of RGs.
Since this aspect is the crux of our investigations, 
we do not consider the K2000 model any further. 

The primary modification made to the other models, 
the incorporation of a growing hotspot size, 
produces the following major results. 
The MBRW model (results in \S\ref{sec:prelim-1D-KS-test}, Table~\ref{tab:1DKS-MBRW}) 
is a substantially better fit to the data than the original BRW model (\S5.2 in BW), 
as the total 1-D K-S probability is better by $\sim 4 \sigma$ in the default case, 
and by $\sim 2.5 \sigma$ in the ``best-fit'' case, of MBRW when compared to BRW. 
The 1-D K-S probabilities for $\alpha$ are sometimes better (especially for 7C) 
and in a few cases (see appendix of \citealt{barai06}) approach the value 0.01 
for which a model spectral index fit is not firmly rejected. 
The MMK model produced fitting statistics 
(\S\ref{sec:prelim-1D-KS-test}, Table~\ref{tab:1DKS-MMK}) 
which are better than or comparable to the original MK model fits (\S5.2 in BW). 

We have explored the modified models through our extensive multi-dimensional 
Monte Carlo simulation procedures and parameter variations in the modified models. 
But similarly to what we found for the original models in \S5 of BW, 
we found that no modified model gives acceptable fits 
to {\it all} the source characteristics, $[P, D, z, \alpha]$, 
for {\it all} the three surveys 3C, 6C and 7C, simultaneously. 

Steepening the power law index for the initial beam power distribution 
(Eq.~2 of BW) to $x=3$ (from $x=2.6$ used by BRW) 
while using the default maximum age $T_{Max} = 500$ Myr, 
improved the 1-D K-S statistics for the MBRW model, 
as can be seen from Table~\ref{tab:1DKS-MBRW}. 
However, in the MMK model $x=2.6$ gave better results, 
as compared to $x=3$ (Table~\ref{tab:1DKS-MMK}). 
This is the only model we examined 
which gave better 1-D K-S fits 
with the beam power distribution index set to $x=2.6$. 
Simulations done by co-varying $T_{Max}$ and $x$ 
gave better fits at a maximum age of 300 Myr (MBRW) and 150 Myr (MMK), 
when combined with the above ``optimal'' values of $x$. 
These ``best-fit'' values of $x$ and $T_{Max}$ for the 
modified models (except $x$ for MMK) are comparable to those 
found earlier for the original models (\S6 of BW).

\subsection{Comparing Models with Additional Statistical Tests} 
\label{sec:model-compare-AddStatTest}

\begin{figure*} [!t] 
\centering
\includegraphics[scale=0.9]{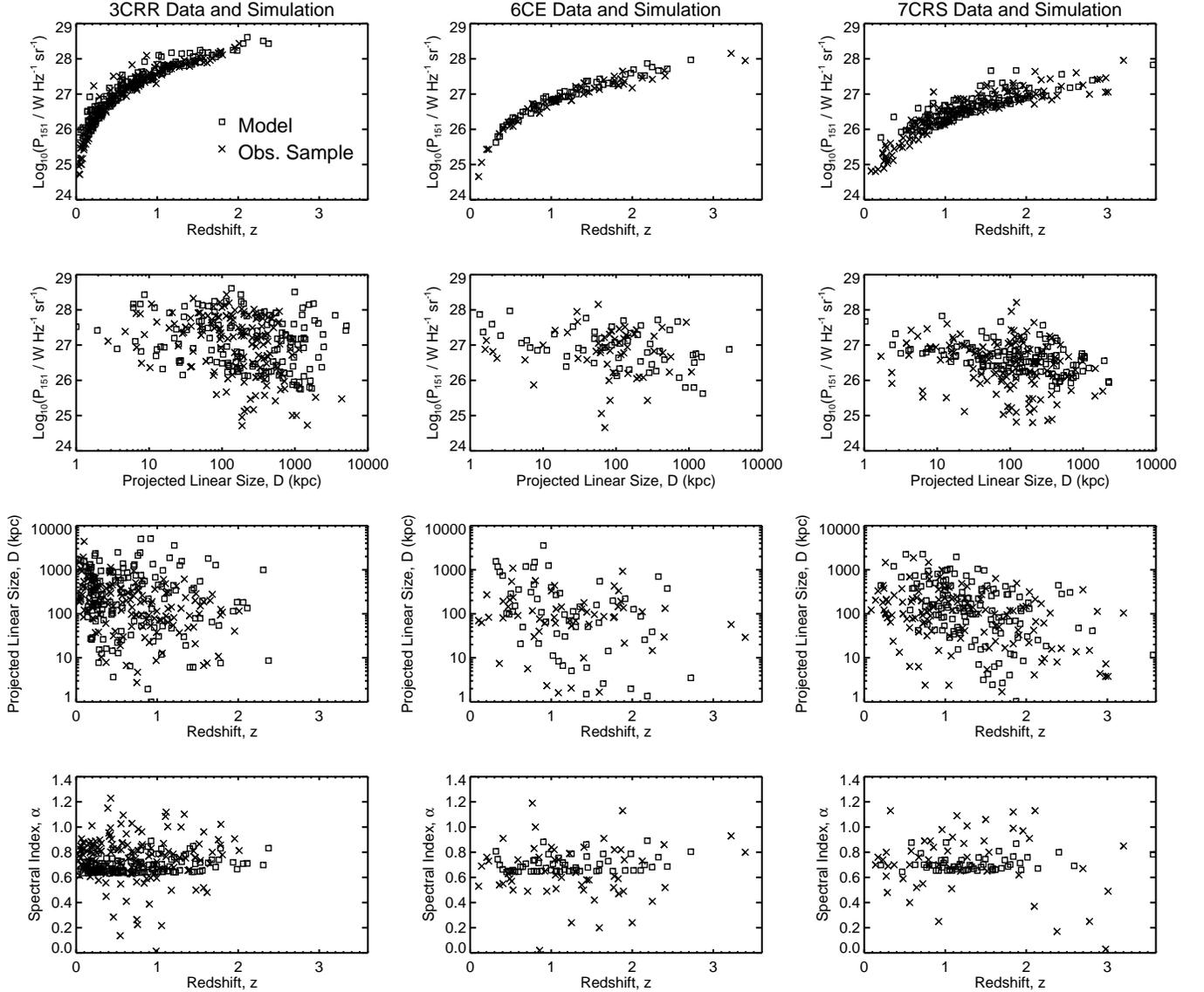}
\caption{The 3CRR, 6CE and 7CRS survey data, overplotted with 
the $[P$--$D$--$z$--$\alpha]$ planes for the 3C, 6C and 7C simulations 
of a good fit of the MBRW model. 
The initial ensemble (of size 4963343) is generated using 
$x = 3.0$, $T_{Max} = 300$ Myr; 
the power evolution is with the parameter variation of $t_{bf} = 100$ yr,
with the rest being their default values as in the MBRW model (\S\ref{sec:ModifiedModels}).
The 1-D K-S statistics for this case are in Table~\ref{tab:1DKS-MBRW} (last but 6th entry). 
} 
\label{fig:BestFitBRWMODallC}
\end{figure*}
Examining the 2-D K-S test results given in Table~\ref{tab:2DKS-all} 
we can say that the $[P$--$z]$, $[P$--$D]$ and $[z$--$D]$ planes can be reasonably fitted 
by the ``varied'' cases of the modified models. 
Six of these 9 planes not involving $\alpha$ (those three slices
for each of the 3C, 6C and 7C comparisons) had  2-D probabilities  $> 0.1$ 
for the MMK ``varied'' model; 
this is true for 4 ${\cal P}$'s in the MBRW cases. 
All of the 2-D ${\cal P}$'s of the ``varied'' MMK model 
are higher than those of the default MMK. 
When compared to the corresponding default versions, 
improvements are seen for 8 of 9 of the 2-D ${\cal P}$'s 
not involving $\alpha$ in the ``varied'' MBRW model. 
These 2-D results provide support for the superiority of the ``varied'' models 
(selected from 1-D K-S tests) in fitting the data. 

Comparing the ``varied'' cases of the two modified models 
themselves in the 9 planes not involving $\alpha$, we see that
6 of the 2-D ${\cal P}$'s for the MMK model are higher than those of MBRW. 
Nonetheless, the $\alpha$-related 2-D K-S probabilities are $\leq 0.008$ for every modified model. 
So, similarly to the original models, 
the modified models cannot fit any plane involving $\alpha$. 
From the 2-D K-S probabilities 
we conclude that the MMK model is better 
(having the highest number of 2-D ${\cal P}$'s close to 1) 
in fitting the observational data than is the MBRW model. 

From the Spearman partial rank correlation analyses 
on the combined data of the 3 surveys 
(Table~\ref{tab:CorrCoeff-all}) 
we conclude that 
many matches to the data correlations are acceptable for the MBRW model, 
but they are less good for the MMK model. 
It is interesting to note that the parameter variation cases which were the best fits 
(i.e., gave the highest combined probability, ${\cal P}_{[P, D, z, \alpha]}$) 
from the 1-D K-S results, 
or the ``varied'' cases in Tables~\ref{tab:2DKS-all} and \ref{tab:CorrCoeff-all}, 
are not necessarily the best fits according to the correlation analyses. 
For the MBRW and MMK models 
the default and the ``varied'' cases perform comparably, 
as 3 correlations are better in the default, and 
the remaining 3 are better in the ``varied'', models.

\begin{figure*} [!t] 
\centering
\includegraphics[scale=0.9]{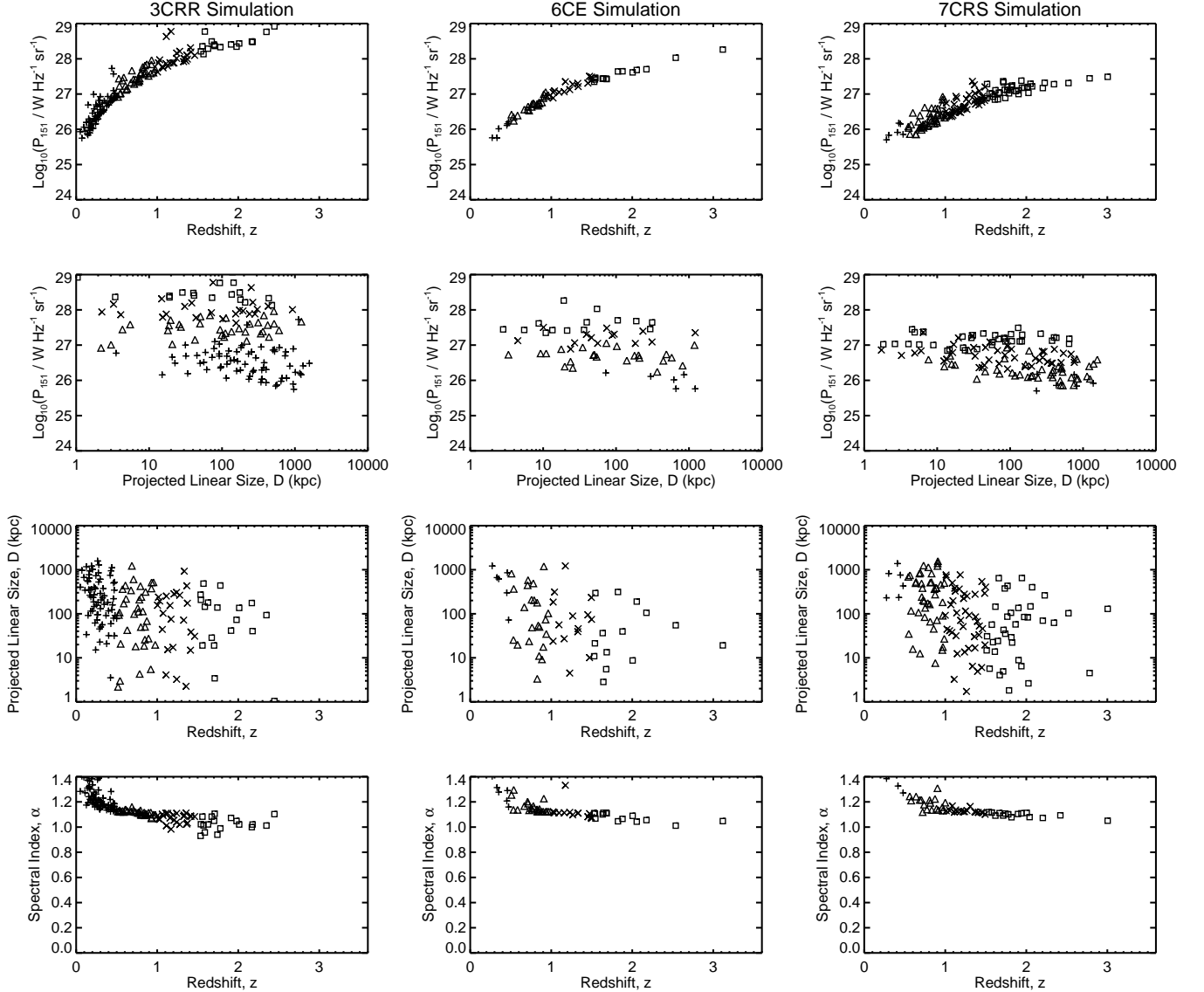}
\caption{The $[P$--$D$--$z$--$\alpha]$ planes for the 3C, 6C and 7C simulations 
of a good fit of the MMK model. 
The initial ensemble (of size 3888492) is generated using 
$x=2.6$, $T_{Max}=150$ Myr; 
the power evolution is with parameter variation $\beta = 1.6$,
the rest of the parameters having their default values of the MMK model (\S\ref{sec:ModifiedModels})). 
The symbols classify the sources into redshift bins as follows; 
{\it Plus}: $0 \leq z < 0.5$, {\it Triangle}: $0.5 \leq z < 1.0$, 
{\it Cross}: $1.0 \leq z < 1.5$, {\it Square}: $1.5 \leq z$. 
The 1-D K-S statistics for this case are in Table~\ref{tab:1DKS-MMK} (9th entry). 
}
\label{fig:BestFitMKMODallC}
\end{figure*}
Considering the signs of the four-variable coefficients of the combined surveys, 
the MMK model predicts a $[P$--$\alpha]$ anti-correlation, a $[D$--$z]$ correlation; 
its ``varied'' case also produces a $[P$--$z]$ anti-correlation. 
All these are trends opposite to those seen in the survey data and to the other models, 
except that the MBRW model also predicts a $[P$--$\alpha]$ anti-correlation. 
Both the MMK models are the only ones which produce the correct signs of the $[D$--$\alpha]$ and $[z$--$\alpha]$ correlations of the 
combined survey data   
(the $[D$--$\alpha]$ correlation is also exhibited weakly by the MBRW default case). 
However, this advantage is not meaningful as the MMK model 
gives very poor fits to the actual $\alpha$-distributions. 

According to the correlation coefficient analyses involving the data 
from all the surveys (3C, 6C and 7C) the 
the MBRW model fits the data most closely, followed by MMK. 
In particular, the MBRW model provides the best fit to the key
4-variable correlation, $r_{PD, z\alpha}$. 
This indicates that in the BRW model a growing hotspot is able to reproduce the 
$P$--$D$ evolution (seen in 3C, 6C and 7C survey data) 
better than assuming a constant hotspot size (the original BRW model).
Similar trends emerged when we examined the 3-variable correlation coefficients. 


\subsection{$[P$--$D$--$z$--$\alpha]$ Planes} 
\label{sec:P-D-z-a-planes} 

We plotted planes through the $[P$--$D$--$z$--$\alpha]$ volume 
for the simulated surveys using the modified models, 
and compared them with the overall trends in the 
$[P$--$D$--$z$--$\alpha]$ slices of the observational data. 
The actual 3C, 6C and 7C data are plotted as crosses and 
the simulated data (3C, 6C and 7C virtual surveys) for the 
``best-fit'' or  ``varied'' parameter set for the MBRW model 
(those which give the highest total 1-D K-S probability) 
are overplotted as squares in Fig.~\ref{fig:BestFitBRWMODallC}. 
The Monte Carlo results for the ``best-fit'' parameter set 
for the MBRW model are shown in Fig.~\ref{fig:BestFitMKMODallC}. 
The main features of the $[P$--$D$--$z$--$\alpha]$ planes of the modified models 
are analogous to those of the original models in \S6 of BW,  
so we discuss them only briefly, placing stress on any new features. 

In the $[P$--$z]$ plane, 
all of our simulated surveys of all the modified models 
miss many of the low-$P$ sources seen in the data. 
Too few low-$z$/low-$P$ sources are produced in all the simulated 7C surveys. 
There is underproduction of very high-$z$ sources ($z>2$) in the 7C simulations 
and a similar, but less pronounced, trend is also present for 6C. 

From the trends in the $[P$--$D]$ plane, 
the MBRW model overproduces large powerful sources in 3C, 
and underproduces the large weaker sources. 
The 6C and 7C $[P$--$D]$ planes of all the modified models show a 
closer match to the data, but the actual data is more scattered
than the simulations. 
Given that any additional physics not included in the models 
would tend to broaden this distribution, this result is expected. 

There is $P$--$D$ anti-correlation in the MMK model in 
all of the 3C, 6C and 7C simulations. 
Such $P$--$D$ evolution is also seen in the MBRW model 
where it is more pronounced in the 6C and 7C simulations. 
An important improvement in the MBRW model is that 
the strong $P$--$D$ anti-correlation apparent in even the
 ``best-fit'' BRW model is diluted 
when modified to incorporate a growing hotspot. 

In the $[D$--$z]$ plane, 
the MBRW model still overproduces the larger 3C sources. 
The MMK model's ``best-fit'' case $[D$--$z]$ planes 
seem to be a good fit to the data (especially the 3C $[D$--$z]$), 
with about the right amount of $D$--$z$ anti-correlation. 
The MBRW model produces a weaker $D$--$z$ anti-correlation than does the data in 3C, 
but one stronger than in the data in 6C and 7C. 
An explanation of the $D$--$z$ evolution as an
effect of the ``youth-redshift degeneracy'' 
was given while discussing the original model planes in \S6 of BW. 
From these trends we conclude that the MMK model is 
the best fit (by eye) to the $[D$--$z]$ planes of the 3C, 6C and 7C data. 

All the other characteristics examined in the simulations show 
tight correlations with the spectral index.
The data shows much greater scatter in $\alpha$ than is seen in any of the models. 
The spectral index distributions of all the model simulations are similar to those in BW. 
Since the $\alpha$ statistical fits are very poor, 
we do not discuss them any further. 

\subsection{Alternative RLF} 
\label{sec:disscussions-GrimesRLF} 

The \citet{grimes04} RLF is tested using different values of $x$ and $T_{Max}$, 
but with the default values of radio lobe power evolution 
model parameters of the different models.
When comparing the simulations using \citet{grimes04} RLF
with respect to those with \citet{willott01}'s (with the same model parameters otherwise),
most of the models give better 1-D K-S statistics
using \citet{willott01}'s RLF.  
The KDA and MK models are better for 5 out of 6 comparisons,
the MBRW model is better for 4 out of 6 total.
The MMK model performs equally using both RLFs.
The BRW is the only model which give better results
using \citet{grimes04} RLF for all 5 comparisons we made. 

So we conclude that the KDA, MK and MBRW models
give better fit to the data using \citet{willott01}'s RLF.
From the runs we performed the MMK model produces comparable results with both the RLFs.
The BRW model is better fit with the \citet{grimes04} RLF,
but in an absolute comparison with the other models (KDA, MK and MBRW),
the fits are still poor.

All the spectral index fits are still very poor using the alternative RLF.
So we can say that altering the redshift birth function
has little effect on the major drawback of these models:
the mismatch of the spectral index behavior between data and simulations.

\section{Relevant Volume Filling Fraction} 
\label{sec:RelVol-Fill-Frac} 

The volume of our ``Relevant Universe'' 
is taken to be the volume of the cosmic baryons which exist as the WHIM 
and have temperatures $10^5 < T < 10^7$ K \citep[e.g.,][]{cen99, cen06, dave01}. 
This warm/hot intergalactic gas appears to contain a very large fraction of the baryons 
($\sim 40 - 50 \%$ by mass) in the universe at the present epoch. 
The WHIM permeates the universe as extended large-scale filamentary structures, 
the junctions of which are the sites of galaxy and cluster formation. 

To calculate the volume fraction of the relevant universe filled by RGs 
we first consider that the initial ensemble of sources generated to yield the 
simulated surveys provides a good estimate of the actual population of RGs 
at different redshifts since, as discussed above, those Monte Carlo simulations do produce decent
fits to the $P$--$D$--$z$ distributions.
Among the millions of RGs generated in such an ensemble, 
only a few tens to hundreds are detected in the simulated surveys. 
The key point is that this is also what happens in reality: 
among the millions of RGs born through the quasar era, 
we can now detect only a few in our flux-limited and redshift-complete
radio surveys. 
Severe energy losses (adiabatic, synchrotron and IC losses) 
conspire to force the vast majority of all RGs to fall below our observational flux limits, 
as discussed in detail in earlier work 
(\citealt*{GKW89, blundell99}; BRW; GKW01; GKW03a,b; GKWO; GKWB04). 
Such power losses are evident in the simulations from the steeply falling 
$[P$--$D]$ tracks of \S\ref{sec:P-D-tracks}. 

Next, the universe is divided into redshift bins (shells), 
and the relevant volume fraction is calculated in each $z$-bin. 
Let the minimum and maximum redshifts of a shell be $z_{min}$ and $z_{max}$. 
The bin-width is taken as 
$\Delta z = 0.02 = \left(z_{max} - z_{min}\right)$. 
The mid-redshift of a bin, $z_{mid} = \left(z_{min} + z_{max}\right)/2$, 
is considered as the epoch of that bin. 
Then, the distribution of the volume fraction is estimated as a function of redshift, 
which is then integrated over the entire quasar era epoch 
to get the total volume contribution of several generations of RGs in the universe. 

\subsection{Volume of WHIM in the Universe}  
\label{sec:Rel-WHIM-Vol} 

We use a consensus flat, dark-energy dominated universe 
\citep{spergel03, spergel06}, 
with Hubble constant $H_0 = 71$ km s$^{-1}$ Mpc$^{-1}$, 
matter density parameter $\Omega_M = 0.3$, 
and vacuum energy density parameter $\Omega_\Lambda = 0.7$. 
The comoving volume over all-sky ($4 \pi$ sr) in a redshift shell 
of the universe between $z_{min}$ and $z_{max}$ is 
(\citealt{hogg99}), 
\begin{equation} 
\Delta V_{co} = \frac{4 \pi}{3} \left(\frac{c}{H_0}\right)^3 \left(\chi_2^3 - \chi_1^3\right) , 
\label{eq:delta-V-Comov} 
\end{equation} 
where, 
\begin{equation} 
\chi_{2,1} = \int_{0}^{z_{max},~z_{min}} 
              \left[ \Omega_M \left(1+z\right)^3 + \Omega_\Lambda \right]^{-1/2} dz. 
\end{equation}

Our big ensemble (initial population) explicitly detects sources over the 3CRR sky survey area, 
$\textrm{Area}_{3C} = 4.23$ sr. 
So the RG population from which these sources are detected lies within a 
smaller comoving volume extending over only the 3CRR sky area, since 
in these simulations the number of sources can be taken as proportional
to the sky area over which they are surveyed. 
If a simulation detects $N_{\rm{sim} (3C)}$ sources,
where there are $N_{\rm{samp} (3C)} = 145$ sources in the real 3CRR survey,
then the 3C detection ratio is written as, 
$\rm{Ratio}_{3C} = N_{\rm{sim} (3C)} / N_{\rm{samp} (3C)}$.
Hence the comoving volume over the effective 3C survey area is,
\begin{equation}
\Delta V_{co, 3C (\rm{eff})} = 
\Delta V_{co} \frac{ \rm{Area}_{3C} \times \rm{Ratio}_{3C} }{4 \pi}.
\end{equation} 
The effective comoving volume of the $z$-shell is then converted to 
the proper volume it had at that epoch, as $\Delta V_{prop}
= \Delta V_{co, 3C (\rm{eff})} / \left(1+z_{mid}\right)^3$. 
The effective relevant volume of the $z$-shell is then the fraction of the
proper cosmological volume of the shell occupied by WHIM.

The WHIM volume fraction, $\xi$, is adopted from the
large-scale cosmological simulations of \citet{cen99}.
\citet{cen06} give an improved WHIM fraction calculation by
explicitly including galactic superwind feedback processes, but 
for our purposes 
there is no significant difference from their previous results.
Hence we obtain the final ``relevant volume of the universe'' inside a $z$-shell, 
$\Delta V_{WHIM} = \xi \Delta V_{prop}$. 

\subsection{Radio Galaxy Volumes and Relevant Fraction} 
\label{sec:RGvol-RelFrac} 

We calculate RG volumes by assuming that the RGs are
cylindrical in shape with total length $D(t)$ (Eq.~4 of BW),
at an age $t$.
The axial ratio, $R_T$ 
gives the ratio of the source length and its width (or diameter). 
The volume occupied by a RG at an age $t$ is,
\begin{equation}
V_{RG} (t) =  \pi \left[\frac{D(t)}{2 R_T}\right]^2 D(t) = \frac{\pi D(t)^3}{4 R_T^2}.
\label{eq:VRG-t}
\end{equation} 
To get a conservative estimate, in all the volume computations 
$R_T = 5$ is used, irrespective of the model (unless otherwise noted). 
This value appears to be an upper bound to the average axial ratio 
based on observations (GKW01 and references therein). 
The difference between this choice of $R_T = 5$ used to calculate the volumes, 
and that in the KDA model, where $R_T = 1.3$ 
gave the best fit to the $[P$--$D$--$z]$ planes, is noteworthy. 
If later work shows that $R_T < 5$ is indeed preferable, 
then a typical radio galaxy volume $V_{RG} (t)$ will be bigger, 
by up to a factor of $\sim 15$, 
thus more strongly favoring the picture of substantial cosmological impact of RGs.

Following the conservative bent of GKW01, we only consider the contributions of the more powerful
FR II RGs despite the fact that the weaker FR I RGs are much more numerous in our
local universe and some of them are also seen to extend for hundreds of kpc.
There are several good reasons for making this approximation, although we cannot
precisely quantify its effect. 
First, the \citet{willott01} RLF indicates that the numbers of
FR I and FR II sources were much more comparable during the quasar era on which we
are focusing, with the lower luminosity sources hardly evolving for $z > 0.7$ while the
higher luminosity sources continue to rise in density with $z$ until $z > 2$.
Second, it is very likely that the efficiency with which jet thrust is converted into
radio flux in FR I's is substantially higher than in FR II's 
\citep*[e.g.,][]{GKW91, baum95}, 
so that a jet of equal power would make a brighter FR I source, thereby skewing the RLF of FR I's upward.
But since the typical FR I has much lower radio flux it is being energized by 
very much weaker jets than those powering a typical FR II.  Hence, on average, the volumes
enclosed by individual FR I's will be substantially lower than those of the FR II RGs considered here.
In addition, since the expansion of FR I's is less rapid they produce weaker (if any) shocks
for shorter times and will have less ability to trigger star formation.

Let the cosmic times of the boundaries of the $z$-shell 
be denoted as $t_{in}$ and $t_{out}$ (corresponding, respectively, to $z_{min}$, $z_{max}$). 
We assume that all the RGs in an ensemble live out to their full lifetime, $T_{Max}$.
So, while the vast majority of FR II radio sources ever born are too faint
to be detected now,
they do expand as long as the AGN is feeding the jets and lobes
and hence contribute to filling part of the universe. 
We count all the RGs in the simulation initial ensemble which have 
any portion of their lives falling in the time range of that $z$-bin.
The volume contribution of all the RGs which are intercepted by a $z$-shell 
are then added to get the total RG volume, $\Delta V_{RG}$. 
The relevant volume fraction in a $z$-shell is
$\Delta \iota (z) = \Delta V_{\rm RG} / \Delta V_{\rm WHIM}$ . 
Integrating $\Delta \iota (z)$ over $z$ gives this volume fraction as,
\begin{equation}
\iota = \int_{0}^{z_{early}} \Delta \iota (z) dz,
\end{equation}
where $z_{early}$ is the earliest redshift of a source in the initial ensemble. 

There is another significant factor which must
taken into account to estimate the total fraction correctly;
this arises from the contributions 
of the several generations of RGs during the quasar era (QE). 
The total volume filled by the 
multiple generations of RGs in the universe over the whole QE 
was roughly taken into account in GKW01. 
They considered the length of the QE as $t_{QE} \sim 2$ Gyr,
and the maximum age of radio sources $T_{Max} = 500$ Myr.
We also adopt this rather large maximum length of  AGN activity in some of our simulations, 
as it was suggested by BRW.  Ages of that order are supported by a variety 
of recent observational studies mentioned in \S 2 of BW;  the most 
recent modeling of RG ages by \citet{machalski06} 
also implies high values for $T_{Max}$. 
GKW01 argued that every place in the universe could have been potentially affected
by $t_{QE}/T_{Max} = 4$ generations of RGs during the entire QE.
So they multiplied the mean of the corrected RLF by $(t_{QE}/T_{Max})$
to get the total proper density
of intrinsically powerful radio sources in the universe. 

In our simulations we obtain the total fraction 
by adding the values of $\Delta \iota (z)$ several times
in intervals of $T_{Max}$ over the entire QE.
The length of the QE is obtained from the temporal length of the epoch
for which $\Delta \iota (z) \geq 5 \%$ of its peak value.
Starting from the high-$z$ end-point of the QE,
values of $\Delta \iota (z)$ are computed at intervals of $T_{Max}$ and summed, 
until the low-$z$ end-point of the QE is reached or exceeded.
This addition is done several times; 
each time the starting point is chosen differently by going back or forward
from the original starting point by integral multiples of $50$ Myr. 
The summed $\Delta \iota (z)$ obtained from these several 
additions (each starting from a different cosmic time),
are then averaged to get the
mean total ``relevant volume fraction'', $\zeta$, of the universe filled by
generations of radio galaxies during the quasar era.

\begin{figure} [!t] 
\centering 
\includegraphics[scale=0.4]{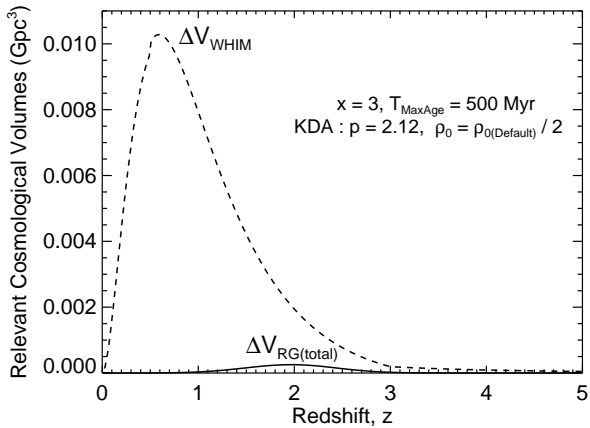} 
\caption{Relevant volumes in the Universe: 
WHIM volume over the effective 3CRR sky area, and total RG volume, 
for the model with parameters noted as a function of redshift.
} 
\label{fig:RelVolUnivRG}
\end{figure}
\subsection{Results and Discussion} 
\label{sec:VolFrac-Results} 

The relevant volume fraction, $\zeta$, was computed for
a subset of the model simulations done with  
the 3 main models, KDA, BRW and MK, 
as well as for the modifications we have considered in the present work.

We now consider the model parameters which determine 
the relevant volume fraction. 
The distribution functions of RG $z$ (Eq.~\ref{eq:rho-z}: $z_0, \sigma_z$),
and of jet power (Eq.~2 of BW: $x, Q_{min}, Q_{max}$),
according to which an initial ensemble of sources are generated 
following the presciption from BRW, 
along with the maximum age, $T_{Max}$, are the parameters which are model-independent in the sense that 
they do not depend on the RG lobe power evolution models.
The RG volume, $V_{RG} (t)$ (Eq.~\ref{eq:VRG-t}) depends on the models through
the linear size, $D(t)$ (Eq.~4 of BW),
which explicitly involves the ambient density parameters (Eq.~3 of BW: $\rho_0, a_0, \beta$). 
The other model-dependent factor is the detection ratio (\S\ref{sec:Rel-WHIM-Vol}) 
which is obtained from the number of sources in the simulated surveys.


\subsubsection{Cosmological Volumes vs. Redshift} 

\begin{figure} [!t]
\centering
\includegraphics[scale=0.35]{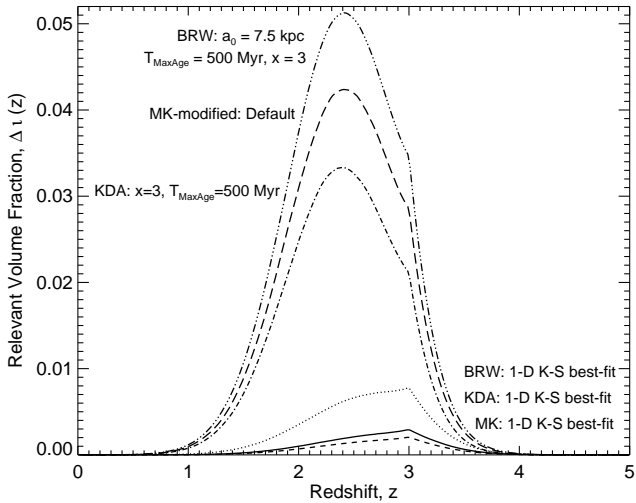}
\caption{Relevant volume fractions of the universe
filled by radio galaxies from several model simulations.
The vertical alignment of the labels correspond to those of the model plots.}
\label{fig:RelVolFracAllModels}
\end{figure}
In Fig.~\ref{fig:RelVolUnivRG}, 
the {\it dashed} curve is the final relevant WHIM volume 
$\Delta V_{WHIM}$ (\S\ref{sec:Rel-WHIM-Vol}), 
and the {\it solid} curve is the total RG volume ($\Delta V_{RG}$, \S\ref{sec:RGvol-RelFrac}), 
for a case of the KDA model.
The RLF or redshift distribution 
from which the sources in the initial ensemble are drawn is a gaussian, 
so the maximum number of sources in the initial ensemble are born near the peak at $z_0 = 2.2$. 
From Fig.~\ref{fig:RelVolUnivRG} we can see that 
the total RG volume, $\Delta V_{RG}$, peaks at $z \sim 2 < z_0$. 
This is because 
the majority of sources born at $z_0$ remain active for $T_{Max} = 500$ Myr,
and are thus counted in several later $z$-bins.
They contribute to the RG volume in increasing amounts as they grow in age,
until the maximum age, $T_{Max}$, 
after which they are assumed to no longer do so.
Their combined increasing contributions at later times
make $\Delta V_{RG}$ peak at a $z < z_0$.
This peak redshift of $\Delta V_{RG}$ should be
around the cosmic epoch corresponding to $t(z_0) + T_{Max}$, 
as that is when the largest number of sources in the population reach their maximum volumes.

On the other hand, Fig.~\ref{fig:RelVolFracAllModels} shows that 
the relevant volume fraction, $\Delta \iota (z)$, peaks at $z > z_0$. 
The distribution of WHIM volume, $\Delta V_{\rm WHIM}$ (Fig.~\ref{fig:RelVolUnivRG}),
can be invoked to explain this result. 
We see that $\Delta V_{\rm WHIM}$ rises sharply from $z \sim 3$, 
until it reaches a peak at $z \sim 0.7$, 
because of the trends of proper volume of $z$-shells in 
the consensus cosmology. 
When the ratio of $\Delta V_{RG}$ to $\Delta V_{\rm WHIM}$
is taken to get $\Delta \iota (z)$ at $z \sim z_0$,
$\Delta V_{RG}$ is divided by a volume $\Delta V_{WHIM}$ which decreases with increasing redshift.

\begin{figure} [!t]
\centering
\includegraphics[scale=0.35]{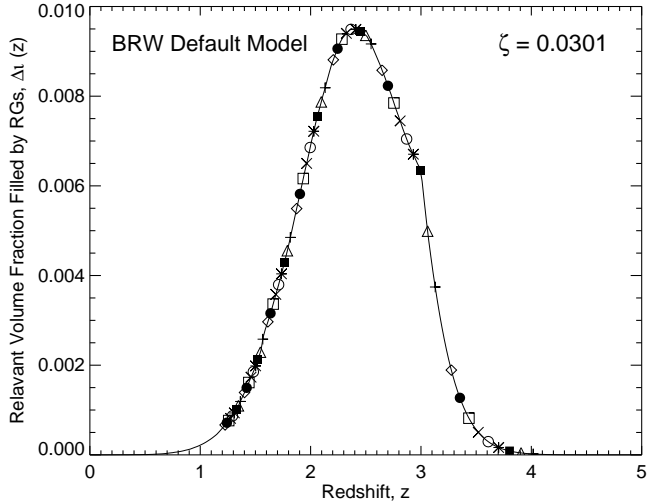}
\caption{Total relevant volume fraction of the universe filled by RGs 
by adding the volume contributions of multiple generations of RGs over the quasar era. 
This is for the BRW simulation with default model parameters, 
for the initial ensemble of size 1561417 
generated using $x = 2.6$, $T_{Max} = 500$ Myr. 
}
\label{fig:AddedVolFracMultGenBRW}
\end{figure}

A representation of how the volume contribution of multiple 
RG generations are added to get the total cumulative fraction
over the entire QE is given in Fig.~\ref{fig:AddedVolFracMultGenBRW}.
The solid black curve in the figure is the volume fraction $\Delta \iota (z)$ 
as a function of redshift.
The symbols (of a single type) plotted on it are the values of $\Delta \iota (z)$
which are picked at intervals of $T_{Max} = 500$ Myr over the QE, and added.
The different plotting symbols denote the different starting points for the added fractions,
which were finally averaged to get the cumulative fraction, $\zeta$. 
For this model of Fig.~\ref{fig:AddedVolFracMultGenBRW} 
(BRW default using initial ensemble with $x=2.6$, $T_{Max} = 500$ Myr), 
the quasar era spans the redshift range $z_{QE} = 3.52 - 1.16$,
or the cosmic time range $1.74 - 5.10$ Gyr,                  
corresponding to a quasar era of duration $t_{QE} = 3.36$ Gyr.  
Hence there are contributions from $\sim 7$ generations 
of RGs in the case where $T_{Max} = 500$ Myr.
The final relevant fraction results for this model are:
\begin{equation}
\iota = 0.0123, ~~~~~ \zeta = 0.0301.
\end{equation} 

\subsubsection{Comparison Between Models} 

Fig.~\ref{fig:RelVolFracAllModels} shows plots of the
relevant volume fraction, $\Delta \iota (z)$, as a function of redshift,
for some of the model simulations. 
Table~\ref{tab:RelVolFrac} gives the relevant volume fraction results 
for the models. 
The results for a particular parameter variation of a model are given in each row. 
Most of the column headings were defined earlier. 
Column 6 gives $A_z$, the normalizing factor in the redshift distribution
(Eq.~\ref{eq:rho-z}) which was used to generate the initial population;
$A_z$ is the factor by which $\left[ V_C (z=0) \times \rho(z_{birth}) \right]$
(paragraph 5, \S2 of BW) is multiplied
to get the number, $N_{born}$, of radio sources born within the relevant comoving volume, $V_C$.
From Table~\ref{tab:RelVolFrac} we see that 
the fractions $\iota$ and $\zeta$ vary significantly both between the models
and for different parameter sets within the same model.
Quite a wide range of relevant volume filling factors can
be produced by the considered ranges in model parameters.

For the ``default'' case, RGs in the BRW simulations 
cumulatively fill $\sim$$3 \%$ of the relevant universe, 
and $\sim$$3.4 \%$ of the relevant volume is filled in the MBRW model. 
This number is $\sim$$4.5 \%$ for the KDA, $\sim$$7 \%$ for the MK 
and $\sim$$13.5 \%$ for the MMK models. 
At the same $x$ and $T_{Max}$, and using default model parameters,
the MK model gives the largest relevant fractions, followed by the KDA and finally by the BRW models.
This is because a larger initial ensemble needs to be generated
in the MK model to yield the same 
number of detected sources compared to the other models. 

Still, the runs involving parameter variations corresponding to
``1-D K-S best-fit''s of the BRW and KDA models give
higher fractions (by $\sim 4 - 5$ times)
than do the corresponding ``default'' models, for the {\it same} $x$ and $T_{Max}$.
This is because the BRW best-fit is with $a_0 = 7.5$ kpc $< a_{0~({\rm Default})}$,
and KDA best-fit is with $\rho_0 = \rho_{0~({\rm Default})}/2$,
both of which have the effect of increasing $D(t)$ and thence the RG volumes.
However, when the favored lower values of $T_{Max}$ are employed, the cumulatively filled
fractions are $0.015 < \zeta < 0.07$.
The relevant volume fractions  for these most preferred 
(with respect to K-S statistics) parameter sets
are all very low compared to the estimate in GKW01, $\zeta = 53 \%$. 
Still, for certain less likely parameter values 
this fraction goes as high as $16 \%$ for BRW, 
$55 \%$ for KDA, 
and $20 \%$ for MK model. 

The relevant fraction is greater for higher $T_{Max}$ (for same $x$),
as is evident from the expression for the RG volume, which scales as $t^{9/(5-\beta)}$.
For the same $T_{Max}$, the fraction is higher with $x=3.0$ than with $x=2.6$.
This might seem counter-intuitive, as with a steeper jet power distribution
the sum of volumes occupied by the same number of sources should be smaller.
The mean RG volume at maximum age, $\langle V(T_{Max}) \rangle$ is indeed smaller
(by a factor of $1.21/1.44 = 0.84$), 
as discussed later (\S\ref{sec:Compare-GKW01}).
Nonetheless, a larger volume fraction at higher $x$ can be explained as 
many more sources must be generated using the steeper slope ($x=3$) to yield
numbers of sources in the simulations comparable to those in the real surveys.
Explicitly, to get the same $\textrm{Ratio}_{3C}$, 
larger ensemble sizes (by $1.5 - 3.5$ times) 
are required for $x=3$ than for $x=2.6$,   
and the increase in ensemble size 
more than offsets the smaller mean volume. 


\subsection{Comparison of Results with a Previous Estimate} 
\label{sec:Compare-GKW01} 

\citet{GKW01} performed a preliminary  calculation to find
the relevant volume fraction added over several generations
of radio sources during the quasar era.
From the results in \S\ref{sec:VolFrac-Results} 
we see that the relevant fraction obtained in this work 
is considerably smaller than the fraction estimated by GKW01. 
For the BRW default model (simulation done with initial ensemble of
size 1561417 generated using $x=2.6$, $T_{Max} = 500$ Myr)
we obtained $\zeta \approx 0.03$ (2nd entry of Table~\ref{tab:RelVolFrac}),
whereas using published graphs in BRW for
the same model GKW01 obtained $\zeta \approx 0.5$. 

There are six differences between the calculations, which taken together, can 
account for most of the discrepancy between the two calculations. 

(1) We adopted the newer consensus cosmology \citep[e.g,][]{spergel06}.
GKW01 used cosmologies with
$H_0 = 50$ km s$^{-1}$ Mpc$^{-1}$, and with either $\Omega_M =$ 0 or 1,
but with $\Omega_\Lambda = 0$.

(2) In the model simulations performed in BW and this work, 
we used $Q_{min} = 5 \times 10^{37}$ W as the minimum jet power, 
following BRW (as described in \S2 of BW).
GKW01 took the effective lower limit of $Q_0$ as $Q_m \equiv 7.5 \times 10^{37}$ W (their \S2.2),
which they inferred (by observing the BRW $[P$--$D]$ tracks in their Figs.~13 and 14)
to be the minimum power a source must have in order to appear in the BRW data set. 
A higher minimum $Q_0$ means that the RG jets will be, on average, more powerful, 
thus making the total RG-volume and the relevant fraction 
higher.

(3) We obtained the average radio galaxy volume at maximum age $T_{Max} = 500$ Myr
to be 1.44 Mpc$^3$, whereas GKW01 obtained a value of 2.1 Mpc$^3$ (their \S2.3),
using the same BRW default model parameters
(except for the different $Q_{min}$ choice). 
Clearly, $\langle V(T_{Max}) \rangle$
should scale as $Q_{min}^{3/\left(5-\beta\right)}$,
since $Q_{max} \gg Q_{min}$.
So GKW01's $\langle V(T_{Max}) \rangle$ should be higher than ours by
$\left( 7.5/5 \right)^{3/(5-\beta)} = 1.42$.

(4) To get the total proper density of RGs, 
GKW01 multiplied the peak of the corrected RLF by
$(t_{QE}/T_{Max}) = 4$ generations of RGs in their \S2.2.
We consider the contribution from multiple generations of RGs by adding the volume fractions $\Delta \iota (z)$ in intervals of $T_{Max}$ over the entire QE (\S\ref{sec:RGvol-RelFrac}),
so we more precisely take into account 
the distribution of $\Delta \iota (z)$ vs.\ $z$.
As can be inferred from Fig.~\ref{fig:AddedVolFracMultGenBRW}, 
simply multiplying the peak fraction 
(highest $\Delta \iota (z)$ in Fig.~\ref{fig:AddedVolFracMultGenBRW}) by $4$,
which was done in GKW01, gives an overestimate of $\zeta$
by a factor of $0.038/0.030 = 1.27$. 

(5) We interpolated the fractional WHIM volume in the universe,
which varies with $z$, from the calculations of \citet{cen99}.
This WHIM volume fraction decreases with increasing redshift,
starting from $\simeq 0.095$ at the present epoch,
to reach $\simeq 0.01$ at $z \simeq 3$.
GKW01 considered a contant WHIM volume fraction of 0.03 
at all redshifts of the QE.
For our computation,
the higher WHIM fraction at low-$z$ dominates over 
the lower WHIM fraction at high-$z$.

(6) In converting from comoving to proper volumes GKW01 used a value of $z = 2.5$.
We integrated over each value of $z$ and had an effective average value of $z \sim 2.2$.
This value is basically the peak, $z_0$, of the
Gaussian redshift distribution (Eq.~\ref{eq:rho-z}) of the radio sources.
This difference causes 
the GKW01 result for $\zeta$ to exceed ours by another
factor of $\sim (1+2.5)^3 / (1+2.2)^3 = 1.31$.


\section{Conclusions and Future Work} 
\label{sec:Conclusions} 

We have performed comprehensive quantitative tests of 
extensively explored modifications to the BRW and MK models for FR II RG evolution 
which allowed the sources' hotspot sizes to grow with age. 
We often found adequate fits to the $[P$--$D$--$z]$ distributions 
for each model for each of the Cambridge catalog subsamples (3CRR, 6CE, 7CRS).
But we cannot locate any parameter sets which provide good simultaneous fits
to all three catalogs and to all four of these observables $[P$--$D$--$z$--$\alpha]$.
Of particular concern are the spectral indices,
where none of the models provides an adequate fit.
We note that all of these 
models only produce a single gross spectral index and that future models must 
improve upon the treatment of the physics that leads to average emissions 
at different frequencies.



From the power vs.\ linear-size tracks 
we see that the BRW $[P$--$D]$ tracks are the steepest among all six models, while
 the MBRW tracks are significantly less steep. 
This causes the most significant change found in the 
model performances 
after modification, 
in that the MBRW model gives substantially better 
K-S statistical fits to the data than the BRW model. 


All the models produced better statistical fits with
the slope of the jet power distribution set to $x = 3$, 
except the MMK model which performed better 
with the default value from BRW, $x = 2.6$.
Considering the active lifetime of the AGN for which the jets feed the lobes, 
we found that 
all the models gave better fits with $T_{Max}$ between 150 -- 300 Myr;
the high default value of 500 Myr is disfavored. 


The KDA, MK, MBRW and MMK models all perform comparably in terms 
of producing high values of total 1-D K-S probabilities. 
From the 2-D K-S test results the ``varied'' cases of most of 
the models can produce adequate fits 
to the $[P$--$z]$, $[P$--$D]$ and $[z$--$D]$ planes. 
Sorting all the models in descending order of the number of non-$\alpha$ 
2-D ${\cal P}$'s greater than any other model 
we have: KDA, MK, MMK, MBRW, BRW. 
From the 4-variable Spearman partial rank correlation coefficient analyses, 
we find that 
the original KDA model can match the survey data correlations 
very closely (at least for $P$, $D$ and $z$), 
followed by BRW and MBRW, then finally the MK and MMK models. 


The existing redshift complete radio source catalogs can limit 
the allowed parameters for each of the models individually and can show that the 
K2000 model is not a good fit to the ensemble of RGs. 
However, they are too small to make a clear distinction among the models we have tested. 
A significant range of the models can provide adequate descriptions 
of the $[P$--$D$--$z]$ distributions. 
Only by obtaining redshift-complete samples covering 
significantly larger portions of the sky than those studied in the 6C and 7C surveys 
would the data have enough power to positively discriminate among the models. 

A major goal of this work has been to 
calculate what fraction of the ``relevant universe'', or 
the large-scale filament-structured WHIM volume, 
do the FR II RGs born over the quasar era cumulatively occupy. 
We found that quite a wide range of relevant volume filling factors can 
be produced, with certain choices of parameters in 
our simulations  producing relevant filling factors as high as $\sim 20\%$ and even $\sim 50\%$; however, 
none of these models  provide good K-S statistical fits to the observations and 
hence these rather high filling factors are unlikely to be realistic.
For both the default and 1-D K-S ``best-fit'' cases of the models using the 
best values of $T_{Max}$ the cumulative volume filling fraction 
of several generations of RGs varied within the range $\sim 1.5 - 7\%$. 
These volume fractions are significantly smaller than the 
preliminary  estimate of GKW01.

We infer that FR II RGs probably cover $\sim 5 \%$ of the WHIM volume cumulatively over the quasar era.
Such filling factors were produced by models providing the best statistical fits to the observations, 
which are quite good fits to the $P$--$D$--$z$ distributions, 
so they are likely to be close to the most probable ``real'' fractions. 
Still, despite the large number of models we have examined they are not
numerous enough to allow a proper error estimate on this key quantity.
The fractional ``internal''  errors on these filling factor values, as obtained 
by  computing results for several of the same models and  same parameter values 
but with different initial random seeds, were in the range  $0.017 < \sigma/\langle \zeta \rangle < 0.10$.



We conclude that the 
expanding radio galaxies born during the quasar era 
can play a significant, though probably not dominant, role in the cosmological history of the universe 
through the triggering of extensive star formation and the spreading of magnetic fields. 


Recently \citet{blundell06} provided observational evidence
for the discovery of low-energy cutoff of particle acceleration
in the lobe of a giant FR II RG.
They obtained a value of $\gamma_{min(hs)} \sim 10^4$ as the
minimum Lorentz factor of particles in the hotspot, 
substantially higher than the values in the 
models where $(1 - 10)$. 
Investigating models using such (tentatively) observationally supported
higher $\gamma_{min(hs)}$ values will be a worthwhile venture. 


In a recent study \citet{kawakatu06} described the
dynamical evolution of the hotspots of radio loud AGN 
using a model significantly different from that of KDA. 
Incorporating such evolving hotspots into the 
models of radio lobe power evolution we analysed 
would be an interesting modification to the models worth exploring.

A potential indicator which can provide an excellent test of whether 
RGs do really trigger galaxy formation, is the 
3-point correlation function between
radio galaxies in large scale galaxy (redshift) surveys \citep[e.g.,][]{borderia91}.
If our RG impact scenario is robust, then there should be a bias 
in this correlation function along the direction of radio lobes of the RGs, 
because more galaxies are formed 
along the radio-axis triggered by jet/lobe expansion, 
as compared to directions perpendicular to the radio-jet. 

The models investigated in this work and in BW predict the power from the radio lobes only.
A natural extension involves the question of
whether the same models also fit deeper radio catalogs
if we take into account the relativistically Doppler boosted core/jet emission. 
By incorporating the beamed core emission,
investigations of simulations of large scale radio surveys
containing many thousands of sources can be done.
Some such deep surveys are
Faint Images of the Radio Sky at Twenty-cm, FIRST \citep{becker95},
the Westerbork Northern Sky Survey, WENSS \citep{rengelink97} and
the NRAO VLA Sky Survey, NVSS \citep{condon98},
which can be made adequately complete in redshift
through optical identifications \citep{ivezic04} from the 
Sloan Digital Sky Survey, SDSS \citep{york00}.
If simulations can be performed to predict thousands of sources,
the possibility of successfully incorporating a
multi-dimensional statistical test becomes much greater.

\acknowledgements 
We thank Christian Kaiser for conversations and clarifying correspondence, 
Konstantina Manolakou for correspondence and providing us with a version of her Fortran code 
and Chris Willott for correspondence and for sending us the 6C and 7C-III data. 
We also benefited from correspondence and conversations with Katherine Blundell, 
Gopal-Krishna, Zeljko Ivezi{\'c} and Steve Rawlings.
We thank the anonymous referee for several useful suggestions which clarified the presentation
of our results.
We are most grateful to Angela Osterman for her efforts on initial versions of some codes, 
PJW is grateful for continuing hospitality at the 
Department of Astrophysical Sciences at Princeton University. 
This work was supported in part by a subcontract to GSU from 
NSF grant AST-0507529 to the University of Washington. 
PB is supported by the Canada Research Chair program and NSERC.


\bibliographystyle{apj} 

\bibliography{apj-jour,PBrefs} 

\clearpage




\begin{deluxetable}{cccccccc} 
\tablewidth{0pc} 
\tablecaption{MBRW Model: 1-D KS Statistics for Selected Parameter Variations\label{tab:1DKS-MBRW}} 
\tablehead{ 
\colhead{$x$} & \colhead{Model} & & & & &  & 
\colhead{${\cal P}_{[P, D, z, \alpha]}$ } \\ 
\colhead{$T_{Max}$\tablenotemark{a}} & 
\colhead{Ensemble Size} & \colhead{Survey} & 
\colhead{${\cal P}(P)$} & \colhead{${\cal P}(D)$} & 
\colhead{${\cal P}(z)$} & \colhead{${\cal P}(\alpha)$} & 
\colhead{${\cal P}_{[P, 2D, z, \alpha]}$} 
} 

\startdata 

2.6 & Default \tablenotemark{b} & 3C & 3.69e-04 & 5.78e-10 & 0.00250 & 6.44e-11 & 0.404 \\* 
500 & 4397469                   & 6C & 0.202    & 4.26e-04 & 0.420   & 3.58e-10 & 0.407 \\* 
    &                           & 7C & 0.00125  & 0.00487  & 0.00317 & 0.00601  &       \\ 
\\ 
2.6 & Default & 3C & 8.42e-06 & 9.61e-08 & 6.86e-04 & 2.80e-11 & 0.451 \\* 
250 & 1466378 & 6C & 0.0709   & 0.00193  & 0.439    & 9.98e-11 & 0.578 \\* 
    &         & 7C & 5.11e-05 & 0.207    & 8.34e-05 & 0.00316  &       \\ 
\\ 
3.0 & Default & 3C & 0.0554  & 4.87e-06 & 0.0990 & 1.11e-15 & 1.06 \\* 
500 & 4886474 & 6C & 0.915   & 0.00708  & 0.404  & 3.66e-10 & 1.12 \\* 
    &         & 7C & 0.00512 & 0.0883   & 0.0117 & 0.00671  &      \\ 
\\ 
3.0 & Default & 3C & 0.152    & 0.00101 & 0.484    & 4.74e-12 & 1.29 \\* 
150 & 3045199 & 6C & 0.434    & 0.00693 & 0.574    & 3.58e-10 & 1.31 \\* 
    &         & 7C & 1.69e-04 & 0.0173  & 2.85e-05 & 7.21e-04 &      \\ 
\\ 
3.0 & Default & 3C & 0.177   & 8.37e-05 & 0.171    & 3.29e-13 & 1.73 \\* 
300 & 4963343 & 6C & 0.481   & 0.0402   & 0.897    & 3.58e-10 & 2.23 \\* 
    &         & 7C & 0.00202 & 0.787    & 4.54e-04 & 0.00296  &      \\ 
\\ 
3.0 & KDA Env. \tablenotemark{c} & 3C & 0.194    & 2.37e-08 & 0.680    & 0.0632   & 1.810 \\* 
300 & $\beta, a_0, \rho_0$       & 6C & 0.520    & 7.50e-04 & 0.956    & 2.33e-07 & 1.812 \\* 
    & 4963343                    & 7C & 7.74e-05 & 0.00214  & 4.28e-05 & 0.0104   &       \\ 
\\ 
3.0 & $\beta = 1.0$ & 3C & 0.0110   & 2.76e-10 & 0.0616   & 3.54e-08 & 0.511 \\* 
300 & 4963343       & 6C & 0.416    & 1.81e-07 & 0.183    & 2.11e-09 & 0.571 \\* 
    &               & 7C & 6.84e-06 & 0.0978   & 2.36e-04 & 0.00112  &       \\ 
\\ 
3.0 & $\beta = 2.0$ & 3C & 0.134    & 1.02e-15 & 0.358    & 0.256    & 0.638 \tablenotemark{g} \\* 
300 & 4963343       & 6C & 0.0908   & 7.07e-06 & 0.335    & 2.72e-04 & 0.639 \tablenotemark{g} \\* 
    &               & 7C & 1.60e-04 & 9.49e-04 & 1.54e-05 & 0.0554   &            \\ 
\\ 
3.0 & $a_0 = 7.5$ kpc & 3C & 0.0530 & 1.01e-07 & 0.0532  & 1.44e-16 & 0.791 \\* 
300 & 4963343         & 6C & 0.362  & 0.0626   & 0.0754  & 1.96e-06 & 1.18  \\* 
    &                 & 7C & 0.0169 & 0.572    & 0.00458 & 0.00861  & \\ 
\\ 
3.0 & $a_0 = 15$ kpc & 3C & 0.197    & 2.25e-04 & 0.477    & 3.01e-08 & 2.07 \\* 
300 & 4963343        & 6C & 0.932    & 0.00148  & 0.408    & 7.24e-09 & 2.62 \\* 
    &                & 7C & 1.32e-04 & 0.893    & 3.99e-04 & 0.00122  &      \\ 
\\ 
3.0 & $a_0 = 20$ kpc & 3C & 0.0621   & 1.42e-04 & 0.112    & 9.34e-07 & 0.901 \\* 
300 & 4963343        & 6C & 0.564    & 0.153    & 0.408    & 2.11e-09 & 1.01  \\* 
    &                & 7C & 9.35e-07 & 0.0247   & 4.28e-05 & 5.45e-06 &       \\ 
\\  
3.0 & $\rho_0=\rho_1$ \tablenotemark{d} & 3C & 0.136  & 2.50e-06 & 0.171  & 1.08e-15 & 0.790 \\* 
300 & 4963343                           & 6C & 0.107  & 0.0212   & 0.0126 & 5.35e-06 & 1.15  \\* 
    &                                   & 7C & 0.0367 & 0.565    & 0.0161 & 3.15e-05 &       \\ 
\\ 
3.0 & $\rho_0=\rho_2$ \tablenotemark{e} & 3C & 0.0852   & 4.89e-06 & 0.387    & 3.26e-10 & 1.77 \\* 
300 & 4963343                           & 6C & 0.724    & 0.0582   & 0.556    & 2.11e-09 & 2.26 \\* 
    &                                   & 7C & 2.36e-04 & 0.741    & 1.35e-04 & 1.46e-05 &      \\ 
\\ 
3.0 & $\rho_0=\rho_3$ \tablenotemark{f} & 3C & 0.0229   & 0.0365 & 0.0616   & 3.22e-11 & 0.656 \\* 
300 & 4963343                           & 6C & 0.416    & 0.0202 & 0.287    & 5.79e-13 & 0.783 \\* 
    &                                   & 7C & 4.48e-07 & 0.128  & 2.35e-05 & 6.46e-07 &       \\ 
\\ 
3.0 & $\Gamma_C=5/3$ & 3C & 0.00283 & 0.00201 & 0.0105 & 0.0361  & 1.13 \tablenotemark{g} \\* 
300 & 4963343        & 6C & 0.987   & 0.00293 & 0.764  & 0.0261  & 1.14 \tablenotemark{g} \\* 
    &                & 7C & 0.698   & 0.00197 & 0.569  & 0.00528 & \\ 
\\ 
3.0 & $\gamma_{min(hs)}=10$ & 3C & 0.253    & 1.36e-04 & 0.579    & 6.02e-11 & 2.31 \\* 
300 & 4963343               & 6C & 0.831    & 0.227    & 0.716    & 7.62e-08 & 2.81 \\* 
    &                       & 7C & 7.70e-05 & 0.584    & 1.35e-04 & 0.00117  &      \\ 
\\ 
3.0 & $\gamma_{max(hs)}=10^{10}$ & 3C & 0.177   & 2.79e-05 & 0.171   & 3.29e-13 & 2.29 \\* 
300 & 4963343                    & 6C & 0.943   & 0.234    & 0.994   & 7.62e-08 & 3.01 \\* 
    &                            & 7C & 0.00115 & 0.932    & 0.00111 & 4.56e-04 &      \\ 
\\ 
3.0 & $\gamma_{max(hs)}=10^{16}$ & 3C & 0.227   & 8.37e-05 & 0.220   & 5.35e-14 & 1.09 \\* 
300 & 4963343                    & 6C & 0.533   & 0.00158  & 0.272   & 7.62e-08 & 1.21 \\* 
    &                            & 7C & 0.00455 & 0.199    & 0.00287 & 0.00468  &      \\ 
\\ 
3.0 & $p = 2.001$ & 3C & 0.00898 & 1.59e-05 & 0.0191  & 3.98e-16 & 0.315 \\* 
300 & 4963343     & 6C & 0.0661  & 0.00158  & 0.00675 & 2.33e-07 & 0.490 \\* 
    &             & 7C & 0.0353  & 0.286    & 0.0344  & 0.0198   &       \\ 
\\ 
3.0 & $p = 2.5$ & 3C & 0.253    & 2.36e-04 & 0.451    & 1.20e-11 & 1.79 \\* 
300 & 4963343   & 6C & 0.724    & 0.0361   & 0.556    & 7.24e-09 & 2.06 \\* 
    &           & 7C & 2.34e-05 & 0.412    & 7.56e-05 & 0.00534  &      \\ 
\\ 
3.0 & $p = 2.999$ & 3C & 0.198    & 1.41e-04 & 0.371    & 5.09e-12 & 1.79 \\* 
300 & 4963343     & 6C & 0.869    & 0.0206   & 0.194    & 7.24e-09 & 2.34 \\* 
    &             & 7C & 8.02e-05 & 0.879    & 1.35e-04 & 0.00468  &      \\ 

\\ 
3.0 & $t_{bs} = 10^3$ yr & 3C & 0.130  & 2.70e-05 & 0.0532 & 2.83e-15 & 0.743 \\* 
300 & 4963343            & 6C & 0.509  & 0.00622  & 0.109  & 7.62e-08 & 0.863 \\* 
    &                    & 7C & 0.0371 & 0.190    & 0.0161 & 0.0192   &       \\ 
\\ 
3.0 & $t_{bs} = 10^7$ yr & 3C & 7.23e-04 & 1.90e-07 & 6.75e-04 & 1.28e-07 & 0.397 \\* 
300 & 4963343            & 6C & 0.389    & 0.0110   & 0.0667   & 2.11e-09 & 0.477 \\* 
    &                    & 7C & 0.0113   & 0.121    & 0.0166   & 1.46e-05 &       \\ 
\\ 
3.0 & $t_{bf} = 0.01$ yr & 3C & 0.0978  & 1.02e-07 & 0.130   & 2.79e-15 & 0.809 \\* 
300 & 4963343            & 6C & 0.641   & 0.00150  & 0.188   & 7.62e-08 & 0.857 \\* 
    &                    & 7C & 0.00481 & 0.0770   & 0.00458 & 0.00500  &       \\ 
\\ 
3.0 & $t_{bf} = 10$ yr & 3C & 0.290    & 8.81e-06 & 0.280    & 2.91e-15 & 2.06 \\* 
300 & 4963343          & 6C & 0.771    & 0.0609   & 0.945    & 7.62e-08 & 2.47 \\* 
    &                  & 7C & 4.03e-04 & 0.603    & 6.70e-04 & 0.00976  &      \\ 
\\ 
3.0 & $t_{bf} = 100$ yr & 3C & 0.253    & 8.37e-05 & 0.622    & 8.34e-15 & 2.66 \\* 
300 & 4963343           & 6C & 0.869    & 0.234    & 0.862    & 7.62e-08 & 3.34 \\* 
    &                   & 7C & 6.75e-04 & 0.879    & 3.99e-04 & 0.00481  &      \\ 
\\ 
3.0 & $t_{bf} = 10^3$ yr & 3C & 0.198    & 2.36e-04 & 0.579    & 6.44e-11 & 2.05 \\* 
300 & 4963343            & 6C & 0.878    & 0.00300  & 0.862    & 7.62e-08 & 2.23 \\* 
    &                    & 7C & 3.96e-04 & 0.286    & 1.35e-04 & 0.00500  &      \\ 
\\ 
3.0 & $t_{bf} = 100$ yr     & 3C & 0.113    & 0.00379 & 0.483    & 2.80e-11 & 2.11 \\* 
300 & $\gamma_{min(hs)}=10$ & 6C & 0.869    & 0.159   & 0.716    & 7.62e-08 & 2.62 \\* 
    & 4963343               & 7C & 3.52e-06 & 0.668   & 4.28e-05 & 0.00917  &      \\ 
\\ 
3.0 & $t_{bf} = 100$ yr & 3C & 0.0624  & 2.66e-06 & 0.247    & 1.76e-06 & 1.58 \\* 
300 & $a_0 = 15$ kpc    & 6C & 0.416   & 0.0992   & 0.556    & 2.11e-09 & 2.23 \\* 
    & 4963343           & 7C & 0.00115 & 0.971    & 7.67e-05 & 5.49e-04 &      \\ 
\\ 
3.0 & $\gamma_{min(hs)}=10$ & 3C & 0.115    & 3.97e-07 & 0.192    & 3.54e-09 & 1.17 \\* 
300 & $a_0 = 15$ kpc        & 6C & 0.564    & 2.50e-05 & 0.408    & 7.24e-09 & 1.42 \\* 
    & 4963343               & 7C & 6.84e-06 & 0.412    & 7.67e-05 & 3.76e-05 &      \\ 
\\ 
3.0 & $t_{bf} = 100$ yr & 3C & 0.147    & 1.07e-07 & 0.192    & 0.0385   & 1.17 \\* 
300 & KDA Env. \tablenotemark{c} ($\beta, a_0, \rho_0$) 
                        & 6C & 0.684    & 0.0626   & 0.556    & 6.89e-07 & 1.22 \\* 
    & 4963343           & 7C & 3.45e-06 & 0.0163   & 1.27e-05 & 0.0103   &      \\ 

\enddata 
\tablenotetext{a}{$T_{Max}$ in units of Myr.} 
\tablenotetext{b}{All other parameters are as in the MBRW model (\S\ref{sec:ModifiedModels}): 
                  following \citet{BRW}, with hotspot size increasing according to \citet{jeyakumar00}.} 

\tablenotetext{c}{Parameters defining the external environment density profile are set to those of 
                  the KDA model: $\beta=1.9, a_0=2$ kpc, $\rho_0=7.2\times10^{-22}$ kg m$^{-3}$.} 

\tablenotetext{d}{$\rho_1 = \rho_{0~({\rm Default})}/2 = 8.35 \times 10^{-24}$ kg m$^{-3}$.} 

\tablenotetext{e}{$\rho_2 = 2 \times \rho_{0~({\rm Default})} = 3.34 \times 10^{-23}$ kg m$^{-3}$.} 

\tablenotetext{f}{$\rho_3 = 4 \times \rho_{0~({\rm Default})} = 6.68 \times 10^{-23}$ kg m$^{-3}$.} 

\tablenotetext{g}{Numbers of sources detected in some of the simulated surveys are considerably 
smaller than in the real surveys, so the 1-D K-S statistic does not hold much significance.} 

\end{deluxetable}
\clearpage 




\begin{deluxetable}{cccccccc} 
\tablewidth{0pc} 
\tablecaption{MMK Model: 1-D KS Statistics for Selected Parameter Variations\label{tab:1DKS-MMK}} 
\tablehead{ 
\colhead{$x$} & \colhead{Model} & & & & &  & 
\colhead{${\cal P}_{[P, D, z, \alpha]}$ } \\ 
\colhead{$T_{Max}$\tablenotemark{a}} & 
\colhead{Ensemble Size} & \colhead{Survey} & 
\colhead{${\cal P}(P)$} & \colhead{${\cal P}(D)$} & 
\colhead{${\cal P}(z)$} & \colhead{${\cal P}(\alpha)$} & 
\colhead{${\cal P}_{[P, 2D, z, \alpha]}$} 
} 

\startdata 

2.6 & Default \tablenotemark{b} & 3C & 0.0931 & 0.00650 & 0.124 & 2.80e-45 & 0.727 \tablenotemark{c} \\* 
500 & 4397469                   & 6C & 0.105  & 0.0283  & 0.575 & 1.82e-21 & 0.767 \tablenotemark{c} \\* 
    &                           & 7C & 0.0356 & 0.0309  & 0.130 & 7.92e-22 &            \\ 
\\ 
2.6 & Default & 3C & 0.0491  & 0.521 & 0.535  & 0        & 2.12 \\* 
150 & 3888492 & 6C & 0.624   & 0.102 & 0.506  & 1.83e-24 & 2.88 \\* 
    &         & 7C & 0.00187 & 0.301 & 0.0491 & 6.13e-21 &      \\ 
\\ 
2.6 & Default & 3C & 9.76e-04 & 0.216    & 0.421  & 0        & 1.64 \\* 
250 & 4195764 & 6C & 0.0935   & 3.44e-04 & 0.681  & 1.83e-24 & 2.36 \\* 
    &         & 7C & 2.38e-05 & 0.827    & 0.0115 & 4.30e-20 &      \\ 
\\ 
2.6 & Default & 3C & 0.00152 & 0.0125 & 0.421  & 0        & 1.72 \\* 
300 & 4342468 & 6C & 0.641   & 0.252  & 0.0709 & 1.83e-24 & 2.34 \\* 
    &         & 7C & 0.0105  & 0.732  & 0.235  & 6.13e-21 &      \\ 
\\ 
3.0 & Default & 3C & 1.21e-07 & 0.0366 & 1.47e-08 & 0        & 0.558 \\* 
150 & 4861474 & 6C & 0.212    & 0.371  & 0.0265   & 2.99e-26 & 0.911 \\* 
    &         & 7C & 0.00341  & 0.134  & 0.0541   & 7.92e-22 &       \\ 
\\ 
3.0 & Default & 3C & 2.30e-04 & 0.195  & 6.52e-05 & 1.58e-31 & 0.410 \tablenotemark{c} \\* 
500 & 4886474 & 6C & 0.154    & 0.0125 & 0.00722  & 8.88e-16 & 0.611 \tablenotemark{c} \\* 
    &         & 7C & 0.145    & 0.128  & 0.0532   & 6.13e-21 &            \\ 

    
\\ 
2.6 & KDA Env. \tablenotemark{d} & 3C & 0.00553  & 0.00253 & 0.485   & 0        & 1.55 \\* 
150 & $\beta, a_0, \rho_0$       & 6C & 0.157    & 0.485   & 0.556   & 1.83e-24 & 2.15 \\* 
    & 3888492                    & 7C & 1.24e-05 & 0.483   & 0.00180 & 2.79e-19 &      \\ 
\\ 
2.6 & $\beta=1.0$ & 3C & 0.00368  & 8.37e-05 & 0.474   & 0        & 1.27 \\* 
150 & 3888492     & 6C & 0.331    & 7.50e-04 & 0.498   & 1.83e-24 & 1.54 \\* 
    &             & 7C & 4.26e-05 & 0.434    & 0.00710 & 8.39e-22 &      \\ 
\\ 
2.6 & $\beta=1.6$ & 3C & 0.121   & 0.709 & 0.535  & 0        & 2.73 \\* 
150 & 3888492     & 6C & 0.724   & 0.103 & 0.408  & 1.83e-24 & 4.04 \\* 
    &             & 7C & 0.00479 & 0.879 & 0.0491 & 4.15e-20 &      \\ 
\\ 
2.6 & $a_0=7.5$ kpc & 3C & 0.0913  & 0.615 & 0.360  & 0        & 1.98 \\* 
150 & 3888492       & 6C & 0.707   & 0.102 & 0.125  & 1.83e-24 & 2.91 \\* 
    &               & 7C & 0.00703 & 0.412 & 0.0689 & 6.13e-21 &      \\ 
\\ 
2.6 & $a_0=20$ kpc & 3C & 0.0253   & 0.802  & 0.192   & 0        & 2.28 \\* 
150 & 3888492      & 6C & 0.564    & 0.0212 & 0.533   & 1.83e-24 & 3.64 \\* 
    &              & 7C & 1.34e-04 & 0.893  & 0.00708 & 8.39e-22 &      \\ 
\\ 
2.6 & $\rho_0=\rho_1$ \tablenotemark{e} & 3C & 0.0921   & 0.0260 & 0.0399 & 0        & 0.987 \\* 
150 & 3888492                           & 6C & 0.229    & 0.0117 & 0.655  & 1.83e-24 & 1.27  \\* 
    &                                   & 7C & 6.56e-04 & 0.412  & 0.0108 & 4.30e-20 &       \\ 
\\ 
2.6 & $\rho_0=\rho_2$ \tablenotemark{f} & 3C & 0.00826  & 0.425    & 0.483   & 0        & 1.58 \\* 
150 & 3888492                           & 6C & 0.293    & 7.50e-04 & 0.556   & 1.83e-24 & 2.13 \\* 
    &                                   & 7C & 7.74e-05 & 0.204    & 0.00291 & 6.13e-21 &      \\ 

\\ 
2.6 & $\gamma_{min(hs)}=7$ & 3C & 0.0921   & 0.0697 & 0.101  & 0        & 1.31 \\* 
150 & 3888492              & 6C & 0.224    & 0.102  & 0.556  & 1.83e-24 & 1.93 \\* 
    &                      & 7C & 2.33e-05 & 0.790  & 0.0108 & 8.39e-22 &      \\ 
\\ 
2.6 & $\gamma_{min(hs)}=100$ & 3C & 8.29e-04 & 2.01e-07 & 0.00485  & 0        & 0.199 \\* 
150 & 3888492                & 6C & 0.0809   & 4.05e-06 & 0.199    & 6.76e-22 & 0.216 \\* 
    &                        & 7C & 9.50e-10 & 0.0267   & 9.25e-07 & 1.74e-18 &       \\ 
\\ 
2.6 & $\gamma_{max(hs)}=3 \times 10^8$ & 3C & 0.0674  & 0.212  & 0.360  & 0        & 1.69 \\* 
150 & 3888492                          & 6C & 0.869   & 0.159  & 0.503  & 1.83e-24 & 2.03 \\* 
    &                                  & 7C & 0.00307 & 0.0465 & 0.0491 & 6.13e-21 &      \\ 

\\ 
2.6 & $p=2.001$ & 3C & 0.0629   & 0.00159  & 0.483   & 2.45e-39 & 1.22 \\* 
150 & 3888492   & 6C & 0.575    & 2.55e-05 & 0.362   & 4.45e-21 & 1.31 \\* 
    &           & 7C & 1.36e-04 & 0.131    & 0.00458 & 6.90e-17 &      \\ 
\\ 
2.6 & $p=2.3$ & 3C & 0.0356  & 0.510 & 0.135  & 0        & 2.02 \\* 
150 & 3888492 & 6C & 0.293   & 0.102 & 0.716  & 1.83e-24 & 3.20 \\* 
    &         & 7C & 0.00108 & 0.999 & 0.0238 & 3.72e-20 &      \\ 

\\ 
2.6 & $\epsilon=0.675$ & 3C & 0.0911   & 0.00159 & 0.227  & 0        & 1.07 \\* 
150 & 3888492          & 6C & 0.564    & 0.00158 & 0.408  & 1.83e-24 & 1.19 \\* 
    &                  & 7C & 6.64e-04 & 0.190   & 0.0161 & 6.13e-21 &      \\ 
\\ 
2.6 & $\epsilon=1.4$ & 3C & 0.122    & 0.619 & 0.568  & 0        & 2.22 \\* 
150 & 3888492        & 6C & 0.479    & 0.349 & 0.408  & 1.35e-23 & 3.18 \\* 
    &                & 7C & 2.44e-04 & 0.199 & 0.0108 & 4.30e-20 &      \\ 

\\ 
2.6 & $\eta=0.2$ & 3C & 0.00922 & 0.0365   & 7.11e-04 & 0        & 1.01 \tablenotemark{c} \\* 
150 & 3888492    & 6C & 0.333   & 7.50e-04 & 0.533    & 1.83e-24 & 1.17 \tablenotemark{c} \\* 
    &            & 7C & 0.0173  & 0.204    & 0.293    & 4.30e-20 &           \\ 
\\ 
2.6 & $\eta=0.6$ & 3C & 0.0175   & 0.532   & 0.192    & 0        & 1.44 \\* 
150 & 3888492    & 6C & 0.416    & 0.00152 & 0.556    & 1.83e-24 & 2.05 \\* 
    &            & 7C & 1.10e-07 & 0.128   & 6.81e-04 & 4.15e-20 &      \\ 

\\ 
2.6 & $\tau=2 \times 10^{-4}$ & 3C & 0.0253   & 4.88e-05 & 0.112    & 0        & 0.680 \\* 
150 & 3888492                 & 6C & 0.564    & 1.65e-07 & 0.267    & 1.40e-23 & 0.696 \\* 
    &                         & 7C & 3.58e-06 & 0.0273   & 4.06e-04 & 1.74e-18 &       \\ 
\\ 
2.6 & $\tau=3 \times 10^{-3}$ & 3C & 0.0497   & 0.167 & 0.135   & 0        & 1.37 \\* 
150 & 3888492                 & 6C & 0.0959   & 0.240 & 0.564   & 1.83e-24 & 2.14 \\* 
    &                         & 7C & 1.34e-04 & 0.732 & 0.00708 & 5.91e-21 &      \\ 

\\ 
2.6 & $\epsilon=1.4$ & 3C & 0.0906  & 0.178 & 0.580  & 0        & 1.98 \\* 
150 & $\beta=1.6$    & 6C & 0.416   & 0.242 & 0.408  & 1.40e-23 & 2.76 \\* 
    & 3888492        & 7C & 0.00110 & 0.732 & 0.0108 & 2.79e-19 &      \\ 
\\ 
2.6 & $\epsilon=1.4$ & 3C & 0.0910  & 0.216 & 0.581  & 0        & 2.04 \\* 
150 & $a_0=7.5$ kpc  & 6C & 0.340   & 0.485 & 0.408  & 1.40e-23 & 2.90 \\* 
    & 3888492        & 7C & 0.00108 & 0.565 & 0.0238 & 1.74e-18 &      \\ 
\\ 
2.6 & $\beta=1.6$   & 3C & 0.207  & 0.216 & 0.535  & 0        & 2.26 \\* 
150 & $a_0=7.5$ kpc & 6C & 0.724  & 0.340 & 0.194  & 1.35e-23 & 3.13 \\* 
    & 3888492       & 7C & 0.0106 & 0.732 & 0.0491 & 4.30e-20 &      \\ 
    
\enddata 
\tablenotetext{a}{$T_{Max}$ in units of Myr.} 
\tablenotetext{b}{All other parameters are as in the MMK model (\S\ref{sec:ModifiedModels}): following \citet{MK}, 
                  with hotspot size increasing according to \citet{jeyakumar00}.} 
\tablenotetext{c}{One cannot be confident of the validity of the K-S statistic as the detected sample 
                  in the simulation is smaller (or, much smaller) than in the actual catalog.} 

\tablenotetext{d}{Parameters defining the external environment density profile are set to those of 
                  the KDA model: $\beta=1.9, a_0=2$ kpc, $\rho_0=7.2\times10^{-22}$ kg m$^{-3}$.} 
\tablenotetext{e}{$\rho_1 = \rho_{0~({\rm Default})}/1.5 = 1.133 \times 10^{-23}$ kg m$^{-3}$.} 
\tablenotetext{f}{$\rho_2 = 2 \times \rho_{0~({\rm Default})} = 3.4 \times 10^{-23}$ kg m$^{-3}$.}

\end{deluxetable}
\clearpage 



\begin{deluxetable}{cccccccc} 
\tablewidth{0pc} 
\tablecaption{K2000 Model: 1-D K-S Statistics for a Few Parameter Variations\label{tab:1DKS-K2000}} 
\tablehead{ 
\colhead{$x$} & \colhead{Model} & & & & &  & 
\colhead{${\cal P}_{[P, D, z, \alpha]}$ } \\ 
\colhead{$T_{Max}$\tablenotemark{a}} & 
\colhead{Ensemble Size} & \colhead{Survey} & 
\colhead{${\cal P}(P)$} & \colhead{${\cal P}(D)$} & 
\colhead{${\cal P}(z)$} & \colhead{${\cal P}(\alpha)$} & 
\colhead{${\cal P}_{[P, 2D, z, \alpha]}$} 
} 

\startdata 

2.6 & Default \tablenotemark{b} & 3C & 5.95e-11 & 4.06e-12 & 8.21e-07 & 3.89e-42 & 0.0127 \\ 
500 & 114900                    & 6C & 0.00110  & 3.59e-09 & 0.0185   & 1.80e-24 & 0.0130 \\ 
    &                           & 7C & 9.89e-17 & 5.39e-04 & 1.24e-09 & 1.91e-13 &        \\ 
\\ 
3.0 & Default & 3C & 8.30e-09 & 0.0151   & 2.69e-07 & 3.89e-42 & 0.152 \\ 
150 & 4861474 & 6C & 3.05e-06 & 2.13e-04 & 0.00264  & 1.80e-24 & 0.302 \\ 
    &         & 7C & 1.04e-19 & 0.221    & 1.48e-13 & 1.91e-13 &       \\ 
\\ 
3.0    & Default & 3C & 8.30e-09 & 0.0200   & 3.53e-06 & 2.00e-41 & 0.241 \\ 
150    & 111072  & 6C & 8.19e-05 & 1.52e-04 & 0.00211  & 4.92e-21 & 0.481 \\ 
       &         & 7C & 4.28e-21 & 0.360    & 9.20e-14 & 2.78e-13 &       \\ 
\enddata 
\tablenotetext{a}{$T_{Max}$ in units of Myr.} 
\tablenotetext{b}{All other dynamical and lobe power evolution model parameters 
are same as in the fiducial model of K2000 \citep{kaiser00}.} 
\end{deluxetable}
\clearpage 



\begin{deluxetable}{cccccccc} 
\tablewidth{0pc} 
\tablecaption{Model 1-D K-S Statistics using \citet*{grimes04} RLF\label{tab:1DKS-GrimesRLF}} 
\tablehead{ 
\colhead{$x$} & \colhead{Model\tablenotemark{a}} & & & & &  & 
\colhead{${\cal P}_{[P, D, z, \alpha]}$ } \\ 
\colhead{$T_{Max}$\tablenotemark{b}} & 
\colhead{Ensemble Size} & \colhead{Survey} & 
\colhead{${\cal P}(P)$} & \colhead{${\cal P}(D)$} & 
\colhead{${\cal P}(z)$} & \colhead{${\cal P}(\alpha)$} & 
\colhead{${\cal P}_{[P, 2D, z, \alpha]}$} 
} 

\startdata 


2.6 & KDA     & 3C & 5.59e-08 & 8.84e-06 & 1.01e-04 & 3.01e-08 & 0.494 \\* 
500 & 3712083 & 6C & 0.140    & 0.00193  & 0.310    & 3.66e-10 & 0.690 \\* 
    &         & 7C & 0.0122   & 0.319    & 0.00202  & 0.0106   &       \\ 
\\ 
3.0 & KDA     & 3C & 0.0281  & 0.0976 & 0.0583  & 3.34e-09 & 0.410 \\* 
100 & 1958652 & 6C & 0.107   & 0.0204 & 0.0397  & 1.96e-06 & 0.633 \\* 
    &         & 7C & 0.00303 & 0.120  & 0.00699 & 0.00230  &       \\ 
\\ 
3.0 & KDA     & 3C & 0.0611 & 0.217 & 0.0836 & 1.28e-11 & 1.14 \\* 
150 & 2747159 & 6C & 0.394  & 0.349 & 0.0735 & 6.89e-07 & 1.82 \\* 
    &         & 7C & 0.0173 & 0.264 & 0.0114 & 3.26e-05 &      \\ 
\\ 
3.0 & KDA     & 3C & 0.133  & 0.0125  & 0.180   & 1.36e-10 & 1.44 \\* 
200 & 3212793 & 6C & 0.416  & 0.00615 & 0.191   & 6.89e-07 & 2.14 \\* 
    &         & 7C & 0.0362 & 0.732   & 0.00728 & 0.0198   &      \\ 
\\ 
3.0 & KDA     & 3C & 0.175   & 0.0253 & 0.387   & 5.04e-14 & 1.39 \\* 
300 & 3813260 & 6C & 0.115   & 0.349  & 0.0245  & 6.89e-07 & 2.12 \\* 
    &         & 7C & 0.00728 & 0.524  & 0.00464 & 0.00500  &      \\ 
\\ 
3.0 & KDA     & 3C & 0.122   & 0.00101 & 0.192  & 1.41e-09 & 1.32 \\* 
500 & 5105485 & 6C & 0.0454  & 0.0968  & 0.0255 & 1.96e-06 & 2.26 \\* 
    &         & 7C & 0.00466 & 0.942   & 0.0165 & 0.00468  &      \\ 

\\ 
2.6 & BRW     & 3C & 4.63e-07 & 5.54e-08 & 3.43e-05 & 8.34e-04 & 0.467 \\* 
500 & 3712083 & 6C & 0.0770   & 0.0265   & 0.212    & 3.66e-10 & 0.751 \\* 
    &         & 7C & 5.04e-05 & 0.438    & 9.03e-05 & 0.00304  &       \\ 
\\
3.0 & BRW     & 3C & 5.84e-06 & 2.27e-21 & 2.04e-05 & 0.00483  & 0.303 \\* 
100 & 1958652 & 6C & 0.177    & 0.00158  & 0.194    & 6.76e-07 & 0.363 \\* 
    &         & 7C & 6.84e-06 & 0.0643   & 1.27e-05 & 0.0104   &       \\ 
\\
3.0 & BRW     & 3C & 3.63e-05 & 5.73e-19 & 6.18e-05 & 0.0328   & 0.629 \\* 
150 & 2747159 & 6C & 0.129    & 0.00158  & 0.287    & 2.33e-07 & 0.959 \\* 
    &         & 7C & 6.67e-04 & 0.353    & 1.36e-04 & 0.0104   &       \\ 
\\
3.0 & BRW     & 3C & 5.69e-05 & 5.00e-18 & 5.93e-05 & 0.0264   & 0.828 \\* 
200 & 3212793 & 6C & 0.177    & 0.00622  & 0.287    & 1.89e-06 & 1.33  \\* 
    &         & 7C & 2.29e-04 & 0.533    & 3.99e-04 & 0.0194   &       \\ 
\\
3.0 & BRW     & 3C & 2.91e-04 & 3.53e-16 & 1.05e-04 & 0.0857   & 0.769 \\* 
300 & 3813260 & 6C & 0.113    & 7.50e-04 & 0.287    & 6.76e-07 & 1.17  \\* 
    &         & 7C & 0.0159   & 0.434    & 0.00180  & 0.0198   &       \\ 
\\
3.0 & BRW     & 3C & 3.00e-04 & 2.00e-21 & 2.94e-04 & 4.23e-04 & 0.694 \\* 
500 & 5105485 & 6C & 0.293    & 6.60e-05 & 0.287    & 2.33e-07 & 1.02  \\* 
    &         & 7C & 7.74e-05 & 0.353    & 7.67e-05 & 0.00500  &       \\ 

\\
2.6 & MK      & 3C & 5.82e-11 & 1.27e-09 & 1.95e-05 & 0        & 0.712 \\* 
500 & 3712083 & 6C & 0.161    & 0.00203  & 0.432    & 1.80e-24 & 1.05  \\* 
    &         & 7C & 1.55e-05 & 0.548    & 0.00126  & 1.58e-15 &       \\ 
\\
3.0 & MK      & 3C & 0.0122  & 0.0187 & 0.148   & 0        & 0.674 \\* 
100 & 1958652 & 6C & 0.564   & 0.156  & 0.0142  & 1.40e-23 & 0.819 \\* 
    &         & 7C & 0.00747 & 0.0323 & 0.00189 & 2.09e-15 &       \\ 
\\
3.0 & MK      & 3C & 0.0494  & 0.125 & 0.458   & 0        & 1.36 \\* 
150 & 2747159 & 6C & 0.396   & 0.336 & 0.0142  & 1.40e-23 & 1.95 \\* 
    &         & 7C & 0.00181 & 0.278 & 0.00484 & 2.16e-15 &      \\ 
\\
3.0 & MK      & 3C & 0.00557 & 0.0253 & 0.222   & 0        & 1.26 \\* 
200 & 3212793 & 6C & 0.798   & 0.323  & 0.0735  & 1.02e-22 & 1.73 \\* 
    &         & 7C & 0.00185 & 0.272  & 0.00471 & 2.09e-15 &      \\ 
\\
3.0 & MK      & 3C & 0.0250 & 0.00247 & 0.247   & 0        & 1.11 \\* 
300 & 3813260 & 6C & 0.498  & 0.159   & 0.0136  & 9.89e-23 & 1.61 \\* 
    &         & 7C & 0.0255 & 0.422   & 0.00479 & 2.03e-15 &      \\ 
\\ 
3.0 & MK      & 3C & 0.00366 & 0.0259 & 0.474    & 0        & 0.897 \\* 
500 & 5105485 & 6C & 0.0441  & 0.0200 & 1.69e-04 & 9.89e-23 & 1.24  \\* 
    &         & 7C & 0.0484  & 0.327  & 0.00471  & 3.94e-16 &       \\ 

\\ 
2.6 & MBRW    & 3C & 9.04e-05 & 8.84e-06 & 0.00288 & 1.52e-08 & 0.50524 \\* 
500 & 3712083 & 6C & 0.564    & 2.55e-05 & 0.189   & 7.62e-08 & 0.50527 \\* 
    &         & 7C & 0.0170   & 5.25e-06 & 0.00703 & 0.0194   &         \\ 
\\ 
3.0 & MBRW    & 3C & 0.0443  & 0.0186 & 0.0320   & 3.26e-10 & 0.435 \\* 
100 & 1958652 & 6C & 0.278   & 0.0984 & 0.118    & 7.62e-08 & 0.543 \\* 
    &         & 7C & 0.00486 & 0.0300 & 6.70e-04 & 1.41e-05 &       \\ 
\\ 
3.0 & MBRW    & 3C & 0.152   & 0.00159 & 0.0616  & 1.95e-14 & 0.619 \\* 
150 & 2747159 & 6C & 0.118   & 0.102   & 0.0255  & 7.62e-08 & 0.930 \\* 
    &         & 7C & 0.00307 & 0.264   & 0.00291 & 3.32e-05 &       \\ 
\\ 
3.0 & MBRW    & 3C & 0.194   & 4.87e-06 & 0.192  & 1.28e-13 & 0.938 \\* 
200 & 3212793 & 6C & 0.727   & 0.0103   & 0.118  & 7.62e-08 & 0.950 \\* 
    &         & 7C & 0.00450 & 0.00595  & 0.0107 & 9.01e-05 &       \\ 
\\ 
3.0 & MBRW    & 3C & 0.253   & 0.00159 & 0.192  & 5.09e-12 & 0.707 \\* 
300 & 3813260 & 6C & 0.276   & 0.0117  & 0.0735 & 7.62e-08 & 0.734 \\* 
    &         & 7C & 0.00481 & 0.0194  & 0.0111 & 0.00500  &       \\ 
\\ 
3.0 & MBRW    & 3C & 0.385  & 1.42e-06 & 0.312  & 9.59e-07 & 0.882 \\* 
500 & 5105485 & 6C & 0.120  & 0.00150  & 0.0255 & 7.62e-08 & 0.923 \\* 
    &         & 7C & 0.0358 & 0.0430   & 0.0160 & 0.00976  &       \\ 

\\ 
2.6 & MMK     & 3C & 0.0681  & 0.0125  & 0.474   & 0        & 1.15 \\* 
500 & 3712083 & 6C & 0.724   & 0.00300 & 0.177   & 1.83e-24 & 1.20 \\* 
    &         & 7C & 0.00451 & 0.0294  & 0.00710 & 4.15e-20 &      \\ 
\\ 
3.0 & MMK     & 3C & 0.0200  & 0.00851 & 0.00903 & 0        & 0.210 \\* 
300 & 3813260 & 6C & 0.107   & 0.154   & 0.00189 & 1.83e-24 & 0.316 \\* 
    &         & 7C & 0.00750 & 0.00102 & 0.00114 & 4.00e-20 &       \\ 
\\ 
3.0 & MMK     & 3C & 0.103   & 0.0696   & 0.0206   & 0        & 0.219 \\* 
500 & 5105485 & 6C & 0.00775 & 1.44e-04 & 1.19e-05 & 1.83e-24 & 0.298 \\* 
    &         & 7C & 0.0115  & 0.00920  & 0.00188  & 5.91e-21 &       \\ 

\enddata  
\tablenotetext{a}{All the dynamical and radio lobe power evolution model parameters
are same as in the default version of the corresponding model (KDA, BRW, MK, MBRW, MMK).} 
\tablenotetext{b}{$T_{Max}$ in units of Myr.} 
\end{deluxetable}
\clearpage 



\begin{deluxetable}{cccccccc}
\tablewidth{0pc}
\tablecaption{2-D K-S Test Results for the Two Modified Models\label{tab:2DKS-all}} 
\tablehead{
\colhead{Model} & & 
\multicolumn{6}{c}{2-D KS Probability, ${\cal P}$(K-S)} 
\\
\colhead{Parameters} &
\colhead{Survey} &
\colhead{${\cal P}(P$--$z)$} &
\colhead{${\cal P}(P$--$D)$} &
\colhead{${\cal P}(z$--$D)$} &
\colhead{${\cal P}(P$--$\alpha)$} &
\colhead{${\cal P}(z$--$\alpha)$} &
\colhead{${\cal P}(D$--$\alpha)$} }

\startdata 

MBRW                      & 3C & 0.00234 & 4.17e-10 & 2.10e-08 & 8.27e-08 & 6.07e-08 & 5.47e-12 \\ 
Default \tablenotemark{a} & 6C & 0.458   & 0.00506  & 0.00667  & 1.46e-06 & 1.48e-06 & 1.53e-04 \\ 
                          & 7C & 0.00309 & 8.54e-04 & 0.00222  & 0.00733  & 0.00648  & 0.0112 \\ 
\\ 
MBRW                     & 3C & 0.299    & 0.00119 & 0.00246 & 5.99e-11 & 3.86e-10 & 4.64e-14 \\ 
Varied \tablenotemark{b} & 6C & 0.781    & 0.434   & 0.496   & 3.21e-06 & 7.85e-06 & 2.76e-04 \\ 
                         & 7C & 1.60e-04 & 0.00297 & 0.00237 & 6.00e-05 & 6.28e-05 & 0.00228  \\ 
\\ 
MMK                       & 3C & 0.0366 & 1.48e-04 & 0.00444 & 2.29e-40 & 2.28e-39 & 1.88e-32 \\ 
Default \tablenotemark{c} & 6C & 0.0656 & 0.0325   & 0.183   & 1.70e-17 & 2.32e-18 & 3.76e-16 \\ 
                          & 7C & 0.0151 & 0.00180  & 0.00355 & 3.53e-15 & 5.25e-14 & 2.88e-13 \\ 
\\ 
MMK                      & 3C & 0.147  & 0.143   & 0.421  & 8.35e-40 & 5.53e-37 & 8.64e-30 \\ 
Varied \tablenotemark{d} & 6C & 0.254  & 0.282   & 0.235  & 1.53e-17 & 2.08e-18 & 2.22e-14 \\ 
                         & 7C & 0.0203 & 0.00627 & 0.0290 & 2.27e-14 & 2.96e-13 & 2.91e-12 \\ 
\enddata 
\tablenotetext{a}{Simulations with the respective parameters of MBRW model as used 
in \S\ref{sec:ModifiedModels}. Initial population generated using $x=2.6$, $T_{Max}=500$ Myr.} 

\tablenotetext{b}{MBRW model simulation using initial population with $x=3.0$, $T_{Max}=300$ Myr. 
The power evolution is with parameter change $t_{bf} = 100$ yr, 
other parameters set to their default values, for a case with 
initial source population size = 4963343 (the last but 6th entry in Table~\ref{tab:1DKS-MBRW}).} 

\tablenotetext{c}{Simulations with the respective parameters of MMK model as used 
in \S\ref{sec:ModifiedModels}. Initial population generated using $x=2.6$, $T_{Max}=500$ Myr.} 

\tablenotetext{d}{MMK model simulation using initial population with $x=2.6$, $T_{Max}=150$ Myr.
The power evolution is with parameter change $\beta = 1.6$, 
other parameters set to their default values, for a case with 
initial source population size = 3888492 (9th entry in Table~\ref{tab:1DKS-MMK}).} 

\end{deluxetable}
\clearpage




\begin{deluxetable}{cccccc}
\tablewidth{0pc}
\tablecaption{4-variable Spearman Partial Rank Correlation Analysis \tablenotemark{a}\label{tab:CorrCoeff-all}}
\tablehead{
&
\colhead{Data} &
\multicolumn{4}{c}{Model (combining all surveys \tablenotemark{a})}
\\
& &
\multicolumn{2}{c}{MBRW} &
\multicolumn{2}{c}{MMK}
\\
\colhead{Coeff.} &
\colhead{All \tablenotemark{a}}     &
\colhead{Default} &
\colhead{Varied \tablenotemark{b}}  &
\colhead{Default} &
\colhead{Varied \tablenotemark{b}} }

\startdata 

$r_{PD, z\alpha}$ \tablenotemark{c}      & 0.0303 & 0.183 & 0.0309 & 0.160 & 0.127 \\
$\Sigma_{PD, z\alpha}$ \tablenotemark{d} & 0.478  & 2.94  & 0.489  & 2.57  & 2.02  \\ 
\\ 
$r_{Pz, D\alpha}$      & 0.716   & 0.667   & 0.649   & 0.0754 & -0.103  \\
$\Sigma_{Pz, D\alpha}$ & 14.2    & 12.8    & 12.2    & 1.20   & -1.64   \\ 
\\ 
$r_{Dz, P\alpha}$      & -0.268  & -0.348  & -0.237  & 0.218  & 0.00189 \\
$\Sigma_{Dz, P\alpha}$ & -4.33   & -5.77   & -3.83   & 3.52   & 0.0300  \\ 
\\ 
$r_{P\alpha, Dz}$      & 0.147   & -0.0163 & -0.0139 & -0.640 & -0.750  \\
$\Sigma_{P\alpha, Dz}$ & 2.33    & -0.259  & -0.219  & -12.1  & -15.4   \\ 
\\ 
$r_{D\alpha, Pz}$      & 0.472   & 0.0649  & -0.275  & 0.605  & 0.346   \\
$\Sigma_{D\alpha, Pz}$ & 8.08    & 1.03    & -4.47   & 11.2   & 5.72    \\ 
\\ 
$r_{z\alpha, PD}$      & -0.0234 & 0.132   & 0.194   & -0.590 & -0.609  \\
$\Sigma_{z\alpha, PD}$ & -0.369  & 2.12    & 3.12    & -10.8  & -11.2   \\ 

\enddata

\tablenotetext{a}{The four observables $P$, $D$, $z$ and $\alpha$ for the 
3C, 6C and 7C III surveys (whether real or simulated), combined together in a single sample.}  

\tablenotetext{b}{The particular parameters used are the same as those in 
Table~\ref{tab:2DKS-all} for each of the MBRW and MMK {\it varied} models.} 

\tablenotetext{c}{Spearman partial rank correlation coefficient between two variables 
$P$ and $D$, when the other two variables $z$ and $\alpha$ are kept fixed.} 

\tablenotetext{d}{Significance level associated with the correlation between $P$ and $D$, 
independent of $z$ and $\alpha$.}

\end{deluxetable}
\clearpage



\begin{deluxetable}{cccccccc}
\tablewidth{0pc}
\tablecaption{Relevant Volume Fractions for Selected Models and Modifications\label{tab:RelVolFrac}}
\tablehead{ 
\colhead{$x$} & 
\colhead{$T_{Max}$} & 
\colhead{Ensemble} &
\colhead{Model} &
\colhead{Ratio$_{3C}$} &
\colhead{$A_z$ ($\times 10^{-8}$)} & 
\colhead{Volume} &
\colhead{Total} 
\\ 
&
\colhead{(Myr)} &
\colhead{Size} &
\colhead{Parameters} & &
\colhead{(Mpc$^{-3}$)} &
\colhead{Fraction, $\iota$} &
\colhead{Vol-Frac, $\zeta$} 
} 
\startdata 

\multicolumn{8}{c}{BRW model} \\ 

2.6 & 250 & 1466378 & Default  & 2.12 & 0.931 & 0.00147 & 0.00623 \\ 
2.6 & 500 & 1561417 & Default  & 1.15 & 0.532 & 0.0123  & 0.0301 \\ 

3.0 & 250 & 1571349 & Default  & 1.33  & 0.997 & 0.00204 & 0.00863 \\ 
3.0 & 250 & 3355926 & $a_0=7.5$ kpc \tablenotemark{a}
                               & 0.869 & 2.13  & 0.00964 & 0.0408 \\ 

3.0 & 500 & 2930490 & Default  & 1.21 & 0.997 & 0.0179 & 0.0437 \\ 
3.0 & 500 & 6451283 & $a_0=7.5$ kpc
                               & 1.03 & 2.19  & 0.0667 & 0.163 \\ 

\\ \multicolumn{8}{c}{MBRW model} \\ 

2.6 & 500 & 4397469 & Default & 2.87 & 1.50 & 0.0139 & 0.0339 \\ 
3.0 & 300 & 4963343 & $t_{bf} = 100$ yr \tablenotemark{b} 
                               & 1.94 & 2.66 & 0.00658 & 0.0240 \\ 

\\ \multicolumn{8}{c}{KDA model} \\ 

2.6 & 150 & 1553389 & Default & 2.17 & 1.60 & 0.000324 & 0.00215 \\ 
2.6 & 500 & 4397469 & Default & 2.18 & 1.50 & 0.0183   & 0.0452 \\ 

3.0 & 150 & 1618248 & Default & 0.703 & 1.66 & 0.000796 & 0.00530 \\ 
3.0 & 150 & 4861474 & $p, \rho_0$ \tablenotemark{c,d} 
                              & 0.993 & 4.99 & 0.00331  & 0.0220 \\ 
3.0 & 500 & 3419466  & Default & 0.559 & 1.16  & 0.0428 & 0.106 \\ 
3.0 & 500 & 6451283  & $p, \rho_0$ \tablenotemark{c}
                               & 0.393 & 2.19 & 0.225 & 0.555 \\ 

\\ \multicolumn{8}{c}{MK model} \\ 

2.6 & 150 & 3888492 & Default & 3.26 & 3.99 & 0.000772 & 0.00512 \\ 
2.6 & 500 & 4397469 & Default & 1.37 & 1.50 & 0.0286   & 0.0699  \\ 

3.0 & 150 & 4861474 & Default & 1.20 & 4.99 & 0.00213  & 0.0141  \\ 
3.0 & 150 & 4861474 & $\gamma_{max(hs)}$ \tablenotemark{e, f}   
                              & 1.11 & 4.99 & 0.00230  & 0.0153  \\ 

3.0 & 500 & 4886474  & Default & 0.421 & 1.66  & 0.0844 & 0.206 \\ 
3.0 & 500 & 4886474  & $\gamma_{max(hs)}$ \tablenotemark{e} 
                               & 0.414 & 1.66 & 0.0858 & 0.209 \\ 

\\ \multicolumn{8}{c}{MMK model} \\ 

2.6 & 150 & 3888492 & $\beta = 1.6$ \tablenotemark{g} & 1.59 & 3.99 & 0.00220 & 0.0146 \\ 
2.6 & 500 & 4397469 & Default                         & 0.710 & 1.50 & 0.0552 & 0.135  \\ 
\enddata

\tablenotetext{a}{1-D K-S best-fit case of BRW.} 
\tablenotetext{b}{1-D K-S best-fit case of MBRW.} 

\tablenotetext{c}{Parameter variations: $p = 2.12$ and
$\rho_0 = \rho_{0~({\rm Default})}/2 = 3.6 \times 10^{-22}$ kg m$^{-3}$.} 
\tablenotetext{d}{1-D K-S best-fit case of KDA.} 

\tablenotetext{e}{Parameter variation: $\gamma_{max(hs)} = 3 \times 10^8$.} 
\tablenotetext{f}{1-D K-S best-fit case of MK.} 
\tablenotetext{g}{1-D K-S best-fit case of MMK.} 

\end{deluxetable}
\clearpage

\end{document}